\documentclass{aa}
\usepackage{txfonts}
\usepackage{graphicx}
\usepackage{natbib}
\bibpunct{(}{)}{;}{a}{}{,}

\begin{document}

\newcommand{\be}{\begin{equation}}
\newcommand{\ee}{\end{equation}}
\newcommand{\bdm}{\begin{displaymath}}
\newcommand{\edm}{\end{displaymath}}
\newcommand{\bea}{\begin{eqnarray}}
\newcommand{\eea}{\end{eqnarray}}

\title{Non-thermal emission from relativistic MHD simulations of pulsar wind nebulae: from synchrotron to inverse Compton}

\author{D. Volpi        \inst{1}
\and    L. Del Zanna    \inst{1}
\and    E. Amato        \inst{2}
\and    N. Bucciantini  \inst{3}
}

\institute{
Dipartimento di Astronomia e Scienza dello Spazio,
Universit\`a di Firenze, Largo E. Fermi 2, 50125 Firenze, Italy
\\ \email{delia@arcetri.astro.it}
\and
INAF - Osservatorio Astrofisico di Arcetri, Largo E. Fermi 5, 50125 Firenze, Italy
\and
Astronomy Department, University of California at Berkeley,
601 Campbell Hall, Berkeley, CA 94720-3411, USA
}

\date{Received ...; accepted ...}

\authorrunning{D. Volpi et al.}
\titlerunning{Non-thermal emission from relativistic MHD simulations of PWNe: from synchrotron to inverse Compton}

\abstract
{}
{The goal of this paper is twofold. In the first place we complete the set of diagnostic tools for synchrotron emitting sources presented by Del Zanna et al. (\emph{Astron. Astrophys. 453, 621, 2006}) with the computation of inverse Compton radiation from the same relativistic particles. Moreover we investigate, for the first time, the gamma-ray emission properties of Pulsar Wind Nebulae in the light of the axisymmetric \emph{jet-torus} scenario.}
{The proposed method consists in evolving the relativistic MHD equations and the maximum energy of the emitting particles including adiabatic and synchrotron losses along streamlines. The particle energy distribution function is split in two components: one corresponding to the radio emitting electrons interpreted as a relic population born at the outburst of the supernova and the other associated with the wind population continuously accelerated at the termination shock and emitting up to the gamma-ray band. The inverse Compton emissivity is calculated using the general Klein-Nishina differential cross-section and three different photon targets for the relativistic particles are considered: the nebular synchrotron photons, photons associated with the far-infrared thermal excess and the cosmic microwave background.}
{When the method is applied to the simulations that better reproduce the optical and X-ray morphology of the Crab Nebula, the overall synchrotron spectrum can only be fitted assuming an excess of injected particles and a steeper power law ($E^{-2.7}$) with respect to previous models. The resulting TeV emission has then the correct shape but is in excess of the data. This is related to the magnetic field structure in the nebula as obtained by the simulations, in particular the field is strongly compressed near the termination shock but with a lower than expected volume average. The \emph{jet-torus} structure is found to be clearly visible in high-resolution gamma-ray synthetic maps too. We also present a preliminary exploration of time variability in the X and gamma-ray bands. We find variations with time-scales of about 2 years in both bands. The variability observed originates from the strongly time-dependent MHD motions inside the nebula.}
{}

\maketitle

\section{Introduction}
\label{sect:intro}
Pulsar wind nebulae (PWNe, or plerions) are a class of supernova remnants (SNR), that originates from the interaction between the ultra-relativistic wind blown by a pulsar (PW) and the surrounding supernova ejecta. The electromagnetic torques acting on a fast spinning, highly magnetized neutron star convert its rotational energy into the acceleration of a cold magnetized wind, expanding at relativistic velocity. The wind velocity needs to be reduced in order to match the boundary condition of non relativistic expansion of the confining supernova remnant. This happens at a termination shock (TS), where the plasma is slowed down and heated, the magnetic field is amplified and the nebular particle spectrum is thought to be produced. If confinement is indeed efficient, a non-negligible fraction of the energy lost by the pulsar, which is invisible as long as it is locked in the cold relativistic wind, may become detectable in the form of non-thermal radiation emitted by the relativistic particles in the nebula.

From the observational point of view, PWNe are characterized by the emission of a very broad band spectrum of non-thermal radiation, typically extending from the radio to the X-ray and even gamma-ray band. The primary emission mechanism is, at most frequencies, synchrotron radiation by the relativistic particles gyrating in the nebular magnetic field. The synchrotron spectrum is cut off at some maximum frequency (typically in the soft gamma-rays) determined by the magnetic field strength and the maximum energy that particles can attain in the nebula. At higher photon energies the dominant emission mechanism is, most likely, inverse Compton (IC) scattering by the same particles interacting with different target photon fields. For recent reviews of theoretical as well as observational aspects of PWNe see \cite{gaensler06,kirk07,bucciantini06a}.

The most recent developments in the study of PWNe are related to high energy observations of these objects. On the one hand, the latest generation of Cherenkov detectors has shown PWNe to be among the highest energy emission sources in the Galaxy \cite[e.g. ][]{aharonian07,gallant07}. On the other hand, detailed spatial mapping at X-ray frequencies, as made possible by Chandra, has shown that the peculiar axisymmetric morphology, known as \emph{jet-torus} structure, initially seen in the prototype of the class, the Crab Nebula \cite[][]{weisskopf00,hester02}, is also present in a number of other objects: the SNR associated to Vela \citep{helfand01,pavlov03} and to PSR B1509-58 \citep{gaensler02}; G0.9+0.1 \citep{gaensler01}; G54.1+0.3 \citep{lu02}; G130.7\,+\,3.1 \citep{slane04}. Relativistic motions and time-varying small-scale features (\emph{wisps} around the TS and filaments in the torus) have also been reported \citep{weisskopf00,hester02,pavlov03}. This paper intends to deal with all these aspects, presenting for the first time a study of the highest energy emission from PWNe in the context of the \emph{jet-torus} scenario, including an investigation of time variability.

The theoretical interpretation of the jet-torus morphology observed in the optical and X-ray bands is based on the idea that the pulsar wind energy flux is anisotropic, depending on latitude above the pulsar rotational equator \citep{bogovalov02,lyubarsky02}. An energy flux that is maximum in the equatorial plane and decreasing towards the polar axis causes the termination shock to be oblate, closer to the pulsar at the poles \citep{komissarov03,delzanna04}. Understanding the formation of a torus in the equatorial plane and of jets along the polar axis is straightforward in this scenario, where the post-shock flows first converge toward the equator and are then diverted along the symmetry axis by magnetic hoop-stresses. Recently, axisymmetric simulations in the relativistic MHD regime (RMHD) have been able to fully confirm this view \citep{komissarov04,delzanna04}.

The validity of this dynamical interpretation of the jet-torus structure was strengthened by the comparison between X-ray observations of the Crab Nebula and non-thermal emission maps built on top of the flow structure resulting from RMHD simulations \citep{delzanna06}. In order to do this, information on the particle spectrum as a function of position in the nebula is needed. This kind of information is not available in a MHD approach, but Del Zanna et al. (2006) showed how the evolution of the particle spectrum in the nebula, including synchrotron and adiabatic losses along streamlines, can be easily accounted for at least in an approximate way within a conservative MHD scheme, once an equation for the evolution of the maximum particle energy is added to the code. The general method proposed in that work allows to compute synchrotron radiation, polarization and spectral index maps based on MHD simulations. The resulting synchrotron surface brightness maps, when compared with the data, showed that the axisymmetric MHD scenario is not only able to account for the general jet-torus morphology, but also to explain a number of finer scale details. Optical and X-ray spectral index maps showed spectral hardening in the torus and softening toward the PWN border also comparable to the observations \citep{veron93,mori04}. Finally, the mildly relativistic nebular flow, with velocities in the range of those observed along the jets and around the TS ($v\sim 0.5-0.8$c), gives rise to features closely resembling the \emph{rings} and the bright \emph{knot} observed in the Crab Nebula, once Doppler boosting is properly taken into account.

 The combination of RMHD simulations and the diagnostic tools for synthetic synchrotron emission has thus proved to be a very powerful investigation technique. Unknown parameters such as the wind magnetization $\sigma$ \citep{kennel84a} or the magnetic field angular distribution can be inferred by comparing the theoretical predictions with observations. In particular, in \cite{delzanna06} the Crab Nebula optical and X-ray morphology were found to be best matched by assuming a latitude-averaged magnetization of $\sigma_\mathrm{eff}=0.02$ and a narrow \emph{striped} wind region of low toroidal magnetic field along the equator (runA). The relativistic electrons produced at the TS were assumed to be a single power law $E^{-(2\alpha+1)}$ with $\alpha=0.6$ as in the model by \cite{kennel84b}. No attempt at modeling the radio emission was made. The resulting synchrotron spectrum was in excess of the data in the X-ray band (by a factor two) and show an unexpected flattening beyond $\approx 3\times 10^{16}$\,Hz. 

The aim of the present work is to further investigate the reasons for this difficulty in reproducing the observed high energy synchrotron spectrum. In particular we consider whether this could be solved through a different choice of the particle distribution function. Contrary to the magnetic field strength, the latter does not affect the dynamics (provided a spectral index higher than $0.5$) and can thus be changed with no consequences on the nebular morphology. However, while the synchrotron emission depends only on the combination of the magnetic field and particle distribution, the IC emission allows to disentangle the contribution of the two. In order to exploit this fact in the present paper we extend the set of diagnostic tools for non-thermal radiation to the IC scattering process, and we apply the method to our simulations (runA) that best matched the Crab Nebula, extending the spectral coverage to the gamma-rays.

As we mentioned the other recent development in the study of PWNe concerns the detection of very high energy gamma-ray emission from these objects. TeV emission has been detected from the Crab Nebula \citep{aharonian04,aharonian06,albert08}; MSH 15\,-\,52 \citep{aharonian05b}; Vela \citep{aharonian06b,enomoto06}; Kookaburra complex \citep{aharonian06c}; HESS J1825\,-\,137 \citep{aharonian06d}; the composite SNR G 0.9+0.1 \citep{aharonian05a}; the two candidates HESS J1357\,-\,645 and HESS J1809\,-\,193 \citep{aharonian07a}. In this context, TeV emission can in principle result from two different processes: either inverse Compton emission involving relativistic electrons that up scatter lower energy photons, or $\pi^0$ decay involving the presence of relativistic protons that produce pions by nuclear collisions. The presence of relativistic protons in PWNe is also suggested by theoretical reasons. In fact, particle acceleration at transverse relativistic shock waves, such as the pulsar wind TS, is difficult to explain through standard acceleration processes. The most successful model so far is based on resonant absorption by electrons and positrons of the relativistic cyclotron radiation produced by ions, that are predicted to be present in the pulsar wind. The model also requires the ions to be energetically dominant in the wind, though very few by number.

The most likely mechanism at the origin of this emission is inverse Compton scattering by the same particles that produce the nebular synchrotron spectrum. These particles can up scatter different targets: CMB photons, synchrotron emitted photons, starlight and possibly FIR photons due to reprocessing of the nebular radiation by dust within the SNR. Studying IC emission is important for two different reasons. First of all, as anticipated, modeling synchrotron and IC radiation at the same time allows to disentangle the information on the magnetic field strength and particle number density which is always combined when considering synchrotron emission alone \citep{gould65,dejager96}. Moreover, a detailed modeling of the IC component is at present the only way of investigating whether the high energy data leave room for an extra contribution of hadronic origin \cite[e.g. ][]{amato03,bednarek03}. Clarifying this point is also important in view of the quest for galactic sources of cosmic rays at energies around the \emph{knee}. 

This paper presents a general method that could be applied to the entire class of PWNe and more generally to all non-thermal sources. In fact, it consists in evolving in time and space the maximum energy of emitting particles, together with the other MHD dynamical variables, while the appropriate simulation set up (e.g. initial and boundary conditions, resolution) and the shape of the distribution function at injection sites are arbitrary and can be both chosen as the most appropriate for the object under investigation. Here, however, our purpose is not to optimize the dynamical free parameters to reproduce specific nebulae, but to apply our emission model to the runA simulation and compare the results to the Crab Nebula data. The Crab Nebula is, in fact, the brightest PWN also at very high energies, and is regarded as a standard candle for gamma-ray observations. This makes it a natural target for new instruments and a wealth of data is already available from ground and space instruments: EGRET \citep{nolan93}, COMPTEL \citep{kuiper01}, HEGRA \citep{aharonian04}, HESS \citep{masterson05,aharonian06}, MAGIC \citep{albert08}. This year GLAST will be launched and so new data will be available from $20$\,MeV to $300$\,GeV.

The existing gamma-ray observations are here used to constrain the parameters of our model. The synthetic emission up to TeV energies is calculated for the first time based on time-dependent 2-D numerical simulations. In particular we assume the configuration corresponding to runA, as reported in \cite{delzanna06}, and following \cite{atoyan96} we consider a more general distribution function of emitting electrons consisting of two families: one corresponding to radio emitting electrons, that can be interpreted as primordial population of particles and the other corresponding to particles continuously accelerated at the termination shock, responsible for the synchrotron spectrum up to the gamma band. The IC emissivity is calculated (in the optically thin regime, appropriate for these objects) using the general Klein-Nishina differential cross-section. The three different photon targets recognized as the most important for the Crab Nebula \citep{aharonian06} are considered: the nebular synchrotron photons, those responsible for the far-infrared thermal excess \citep{marsden84} and the cosmic microwave background. Previous work applied similar emission recipes only to stationary and radially symmetric models \cite[e.g. ][]{gould79}, and noticeably to the Kennel \& Coroniti \cite[KC: ][]{kennel84a} RMHD model \citep{dejager92,atoyan96}.

The paper is structured as follows. In Sect.~\ref{sect:model} our non-thermal emission model is presented. The integrated spectra as arising from both the KC model (used here to test the emission recipes) and our simulations are shown in Sect.~\ref{sect:spectra} and compared with observations of the Crab Nebula. In Sect.~\ref{sect:maps} we produce surface brightness maps at different energies in the gamma-rays to compare with existing and future images. Finally, time variability is studied in Sect.~\ref{sect:variability}, linking the synthetic X and gamma-ray emission.

\section{The non-thermal emission model}
\label{sect:model}
In this section the formulae used to compute synchrotron and IC emission are explained in detail. As mentioned in Sect.~\ref{sect:intro}, the evolution of the distribution function and the synchrotron emission recipes are adapted from \cite{delzanna06}, while the choice of the distribution function at the TS follows \cite{atoyan96}. Some approximations are necessary to avoid using formulae in which some terms are unknown: for instance it is not possible, within the present scheme, to associate the local pressure value to the corresponding one (along streamlines) at the termination shock. This would require evolving in space and time the entire distribution function, a task which is beyond the goal of this paper. As we will see a few additional approximations are used with the aim of reducing the computational costs, after checking that they do not change significantly the final results.

\subsection{Relativistic electrons}
\label{sect:electrons}
In the Crab Nebula spectrum two main breaks appear: one in the IR and the other in the UV. In order to account for both breaks we consider the presence of two distinct populations of relativistic particles (electrons and positrons). In analogy with previous MHD models we assume the break in the UV to be due to synchrotron cooling of a population continuously accelerated at the TS and responsible for the optical and higher frequency radiation \citep{kennel84b}. We then describe the radio emitting electrons as a different population, which has sometimes been interpreted as primordial \cite[e.g. ][]{atoyan96}, born at the outburst of the supernova. The distribution function of both populations is shaped following \cite{atoyan96}.

Assuming isotropy, the radio emitting particle distribution function per unit solid angle is taken as 
\be
\label{eq:func_distr_r}
f_{\mathrm{r}}(\epsilon)=\frac{A_\mathrm{r}}{4\mathrm{\pi}}\epsilon^{-(2\alpha_{\mathrm{r}}+1)}\exp\,(-\epsilon/\epsilon^{*}_{\mathrm{r}}),
\ee
assumed to be homogeneous in the PWN and constant in time. Here $\epsilon$ is the normalized energy of the particle (its Lorentz factor), $A_{\mathrm{r}}$ is the normalization constant, $\alpha_{\mathrm{r}}$ is the radio spectral index, $\epsilon^{*}_{\mathrm{r}}$ is the radio cut-off energy, which is the energy corresponding to a synchrotron cooling time comparable to the age of the nebula. 

The second population is composed by the relativistic wind particles continuously accelerated at the termination shock and injected downstream into the nebula, with the following distribution function
\be
\label{eq:func_distr_w0}
f_{0}(\epsilon_{0})=\frac{A_0}{4\mathrm{\pi}}(\bar{\epsilon}+\epsilon_{0})^{-(2\alpha_\mathrm{w}+1)}\exp\,(-\epsilon_{0}/\epsilon^{*}_{\mathrm{w}}).
\ee
This has to be evolved along streamlines in the PWN and may change with time. The variables with the index $0$ are calculated at the TS. As for the previous case $A_0$ is a constant to be determined based on the local number density or pressure, $\alpha_\mathrm{w}$ is the wind spectral index and $\epsilon^{*}_{\mathrm{w}}$ is the cut-off energy (corresponding to the maximum synchrotron frequency in the gamma-rays). The additional constant $\bar{\epsilon}$ represents the minimum energy of the wind population, and will be chosen to match smoothly the infrared-optical synchrotron emission of the two families of electrons. 

As far as the evolution of the wind population is concerned, following \cite{delzanna06}, adiabatic and synchrotron losses are taken into account by defining the evolved distribution as 
\be
\label{eq:func_distr_w}
f_{\mathrm{w}}(\epsilon)=\left(\frac{\rho}{\rho_0}\right)^{4/3}\left(\frac{\epsilon_{0}}{\epsilon}\right)^{2}\!f_{0}(\epsilon_{0}),
\ee
where $\rho$ is the rest mass density, thus:
\be
\label{eq:func_distr_w_approx}
f_{\mathrm{w}}(\epsilon)=\frac{A_{\mathrm{w}}}{4\mathrm{\pi}}\,p\,\left(\frac{\epsilon_{0}}{\epsilon}\right)^{2}\!(\bar{\epsilon}+\epsilon_{0})^{-(2\alpha_{\mathrm{w}}+1)}\exp\,(-\epsilon_{0}/\epsilon^{*}_{\mathrm{w}}).
\ee
Here we have used the conservation of particles' number along streamlines and assumed adiabaticity for the pressure ($p\propto \rho^{4/3}$). $A_0=A_\mathrm{w} p_0$, and we take into account synchrotron losses through the quantity $\epsilon_\infty$, which is the maximum possible particle energy at a position along the streamline:
\be
\label{eq:eps0}
\epsilon_0=\frac{\epsilon}{1-\epsilon/\epsilon_\infty}.
\ee
The variables $\rho$, $p$ and $\epsilon_\infty$ are in general functions of space and time and they will be here provided by our code. Notice that adiabatic losses are taken into account only through the pressure term in Eq.~(\ref{eq:func_distr_w_approx}), while we are forced to neglect them elsewhere (terms $(p/p_{0})^{1/4}$ are approximated to $1$), as explained in \cite{delzanna06}.

\subsection{Synchrotron emission}
\label{sect:sync}
The synchrotron spectral power emitted by a single ultra-relativistic particle is
\be
\mathcal{P}^{\mathrm{SYN}}_{\nu} (\nu,\epsilon) = 2\sigma_{\mathrm{T}}c \frac{B_{\perp}^{2}}{8\mathrm{\pi}}\epsilon^{2}\,\delta (\nu-\nu_{\mathrm{m}}),
\ee
where $\sigma_{\mathrm{T}}$ is the Thomson cross-section, $B_\perp$ is the local magnetic field component normal to the particle's velocity and $\nu$ is the observed frequency. Notice that for simplicity we have also considered the monochromatic approximation, where $\nu_\mathrm{m}=0.29 \nu_{\mathrm{c}}$ is the location of the maximum emission \cite[see e.g. ][]{rybicki79} and the critical frequency is given by
\be
\nu_{\mathrm{c}}=\frac{3e}{4\mathrm{\pi} mc} B_{\perp} \epsilon^{2}.
\ee
The emission coefficient is, in the general case
\be
j_{\nu}^{\mathrm{SYN}}(\nu)=\int \mathcal{P}^{\mathrm{SYN}}_{\nu} (\nu,\epsilon)f(\epsilon)\mathrm{d}\epsilon,
\ee
whatever the distribution function $f(\epsilon)$. Thanks to the monochromatic approximation we can readily perform the above integration and find
\be \label{eq:emis_sync}
j^{\mathrm{SYN}}_{\nu}(\nu)= 2\sigma_{\mathrm{T}}c\frac{[\epsilon (\nu)]^{3}}{2\nu}\frac{B_{\perp}^{2}}{8\mathrm{\pi}}f[\epsilon (\nu)],
\ee
where 
\be
\epsilon (\nu) = \left(0.29 \frac{3e}{4\mathrm{\pi} mc} B_{\perp} \right)^{-1/2}\nu^{1/2}.
\ee
It is easy to verify that, in the power law regions of both the distribution functions described in the previous section, the emission coefficient behaves as $j_{\nu}\propto \nu^{-\alpha}$, as expected.

Due to relativistic beaming, the magnetic field component in the above expressions can be written as $B_{\perp}= |\mathbf{n}\times\mathbf{B} |$, where $\mathbf{n}$ is the line of sight direction. To reduce computational costs, here we will consider an isotropically distributed field (discrepancies in the integrated fluxes are less than $3\%$ at $1$\,keV), for which $B_{\perp}=\sqrt{2/3}B/\gamma$, $B$ is the toroidal magnetic field as measured in the laboratory frame and $\gamma$ is the Lorentz factor of the fluid element. The synchrotron spectrum, expressed in terms of the total luminosity per unit frequency, is
\be
\label{eq:sync_lum}
L^{\mathrm{SYN}}_{\nu} (\nu)=\int 4\mathrm{\pi} j^{\mathrm{SYN}}_{\nu} (\mathbf{r},\nu)\mathrm{d}V
\ee 
where the integral is over the nebular volume.

Synthetic surface brightness maps are derived instead by integrating the emissivity along the line of sight alone. The Doppler boosting towards the observer is taken in account as in the paper by \cite{delzanna06}.

\subsection{Inverse Compton emission}
\label{sect:IC}

The Inverse Compton spectral power emitted by a single ultra-relativistic particle with normalized energy $\epsilon$ may be written as
\be
\label{eq:ic_power}
\mathcal{P}^{\mathrm{IC}}_{\nu} (\nu,\epsilon) = c\,h\nu \! \int \left(\frac{\mathrm{d}\sigma}{\mathrm{d}\nu_{\mathrm{t}}}\right)_{\mathrm{IC}}\!\!(\nu_{\mathrm{t}},\nu,\epsilon)\,n_{\nu}(\nu_{\mathrm{t}})\,\mathrm{d}\nu_{\mathrm{t}},
\ee
where $n_{\nu}(\nu_{\mathrm{t}})$ is the number density of target photons per unit frequency $\nu_{\mathrm{t}}$. The general form of the differential cross-section for IC scattering per unit frequency, including both the Thomson and Klein-Nishina regimes, is given by \cite{jones68, blumenthal70}: 
\be
\left(\frac{\mathrm{d}\sigma}{\mathrm{d}\nu_{\mathrm{t}}}\right)_{\mathrm{IC}}\!\!=\frac{3}{4}\frac{\sigma_\mathrm{T}}{\epsilon^2\nu_\mathrm{t}}\left[2q\ln q +(1-q)\left(1+2q+\frac{1}{2}\frac{(\Gamma q)^{2}}{1+\Gamma q}\right)\right],
\ee
with
\be
q=\frac{h\nu / mc^{2}}{\Gamma (\epsilon - h\nu / mc^{2})}~~~\mathrm{and}~~~\Gamma = 4 \epsilon h\nu_{\mathrm{t}}/mc^{2}.
\ee
Due to the kinematics of the scattering, $q$ is a parameter in the range $0<q\le 1$ and $\Gamma$ determines the energy domain (the Thomson limit corresponds to $\Gamma\ll 1$).

As in the previous case, the emission coefficient is given by
\be
j^{\mathrm{IC}}_{\nu}(\nu)=\int \mathcal{P}^{\mathrm{IC}}_{\nu} (\nu,\epsilon)f(\epsilon)\mathrm{d}\epsilon,
\ee
for any distribution function of scattering particles $f(\epsilon)$. However, in the most general IC scattering case we cannot reduce the above formula to an analytical expression and a double integral has to be computed numerically. Following \cite{blumenthal70}, we replace the integration over $\epsilon$ by an integration over $q$. The relation between these two variables is
\be
\epsilon (q)=\frac{1}{2}\frac{h\nu}{mc^{2}}\left[1+\left(\frac{1+sq}{sq}\right)^{1/2}\right],~~~s=\frac{h\nu}{mc^{2}}\frac{h\nu_{t}}{mc^{2}},
\ee
where the dimensionless quantity $s$ is small in the Thomson limit and large in the extreme Klein-Nishina regime.

The target photon density in Eq.~(\ref{eq:ic_power}) may have different origins. Here we study three cases, recognized to be relevant for the Crab Nebula and for PWNe in general \citep{atoyan96}: IC scattering of reprocessed synchrotron photons (IC-SYN), IC scattering of the far-infrared thermal radiation by local dust (IC-FIR), and IC scattering of cosmic microwave background photons (IC-CMB). 

In the self-synchrotron emission case, the photon density at any position $\mathbf{r}$ can be determined from the radiative transfer equation in the optically thin regime \citep{gould79, dejager92, atoyan96}:
\be
n^{\mathrm{IC-SYN}}_{\nu} (\mathbf{r},\nu_{t})=\frac{1}{c\,h\nu_t}\int\frac{j^{\mathrm{SYN}}_{\nu} (\mathbf{r}^\prime,\nu_t)}{|\mathbf{r}^{\prime}-\mathbf{r}|^2}\mathrm{d}V^{\prime},
\ee
where the spatial integral is over the whole nebula. However, in order to reduce the computational costs, after testing that differences are negligible, we use here the approximation of spherical symmetry and of a homogeneous synchrotron emission coefficient. We thus replace the above formula with 
\be
\label{eq:phot_dens_IC}
n^{\mathrm{IC-SYN}}_{\nu} (r,\nu_{t})=\frac{1}{c\,h\nu_{t}}\frac{L^{\mathrm{SYN}}_{\nu} (\nu_{t})}{4\mathrm{\pi} R^2} U(r/R),
\ee
where the synchrotron spectral luminosity is given in Eq.~(\ref{eq:sync_lum}), $R$ is the radius of the nebula and
\be
\label{eq:u_x}
U(x)=\frac{3}{2}\int^{1}_{0}\frac{y}{x}\ln{\frac{x+y}{\vert x-y\vert}}\mathrm{d}y,
\ee
with $x=r/R$ and $y=r^{'}/R$. $U(x)$ is a non-dimensional quantity decreasing with $x$ from $3$ to $1.5$ \citep{atoyan96}.

When the emission recipes are applied to the results of our simulations, the synchrotron photon density in Eq.~(\ref{eq:phot_dens_IC}) becomes a function of time. Variations of the synchrotron emitted flux are up to 30\% on time-scales of order a few years, as we will discuss later on. Since this time-scale is not much larger than the light crossing time of the inner nebula, the temporal lag between the emission of a photon and its comptonization should be taken into account. This could properly be done only by using more complex, time dependent radiation transfer equations, that should be solved simultaneously with the MHD dynamics. Such an analysis is beyond the goal of the present paper, where we will make the approximation of instantaneous comptonization. This implies that our results on gamma-ray variability at frequencies involving a time-varying photon target density should be taken with caution.

In the case of the thermal FIR emission, the emitting dust is expected to be located in a torus created by the supernova progenitor and in the optical filaments \citep{green04}. Since the dust volume and location are uncertain, we choose the crude approximation:
\be
\label{eq:phot_dens_FIR}
n^{\mathrm{IC-FIR}}_{\nu} (\nu_{t})=\frac{1}{c\,h\nu_{t}}\frac{L^{\mathrm{FIR}}_{\nu} (\nu_{t})}{4\mathrm{\pi} R^2},
\ee
with $L^{\mathrm{FIR}}_{\nu}$ corresponding to a uniformly emitting region of black-body radiation. 

Finally, the CMB photon density is provided by the Plank distribution
\be
n^\mathrm{IC-CMB}_{\nu} (\nu_{t})=\frac{8\mathrm{\pi}}{c^{3}}\frac{\nu_{t}^{2}}{\exp(h\nu_t/k_{\mathrm{B}}T)-1},
\ee
where $k_\mathrm{B}$ is the Boltzmann constant and $T=2.7$\,K.

Once the IC emission coefficient has been calculated as a function of space, integration over the PWN volume gives the total spectral luminosity as in Eq.~(\ref{eq:sync_lum})
\be
L^{\mathrm{IC}}_{\nu} (\nu)=\int 4\mathrm{\pi} j^{\mathrm{IC}}_{\nu} (\mathbf{r},\nu)\mathrm{d}V,
\ee
where scattering with different targets will be computed separately to investigate the relative contribution to the overall IC emission.

The surface brightness maps with the relativistic corrections will be computed as for synchrotron emission.

\section{Integrated spectra: fitting the Crab Nebula data}
\label{sect:spectra}

\begin{figure*}
\centering
\resizebox{\hsize}{!}{
\includegraphics{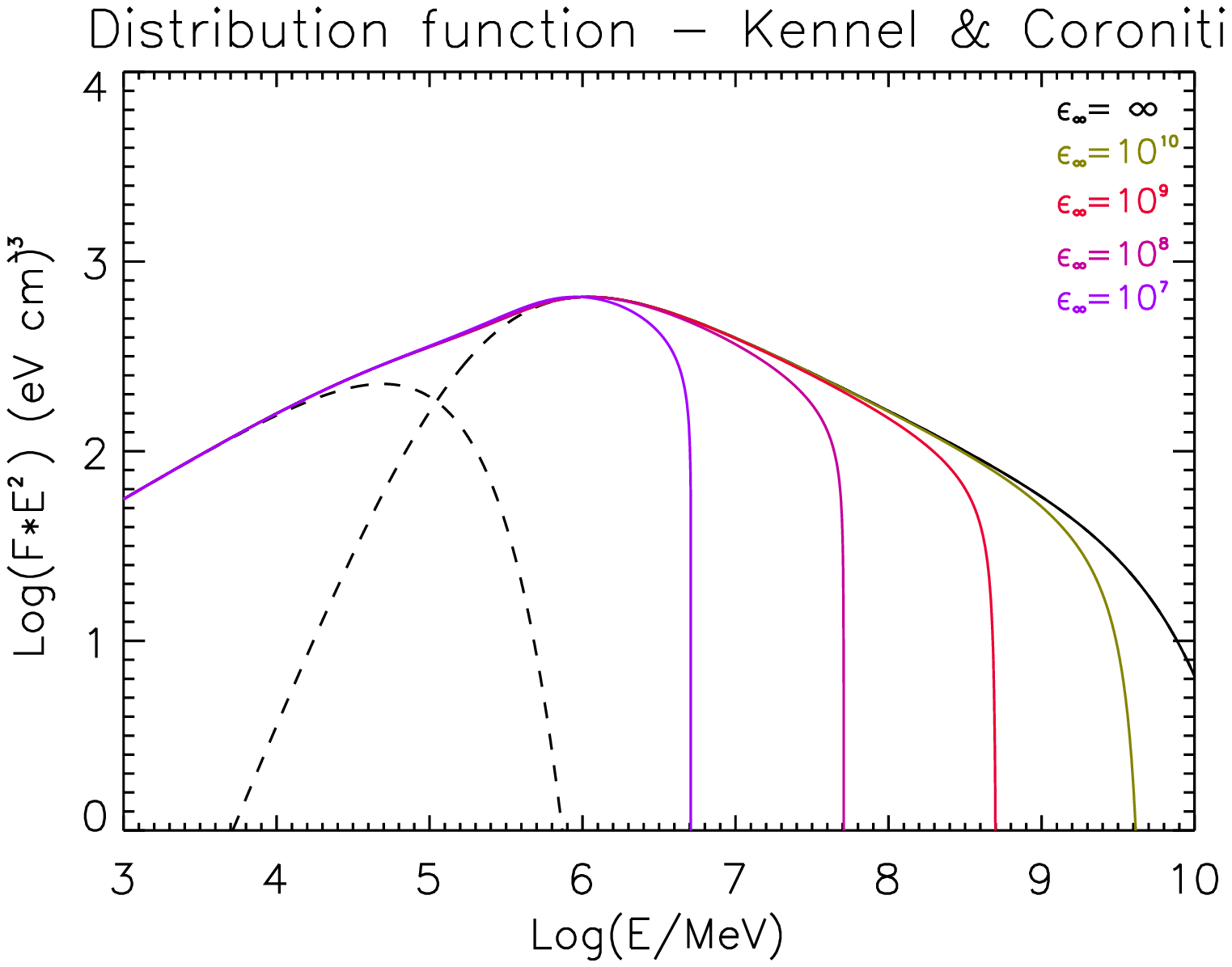}
 \includegraphics{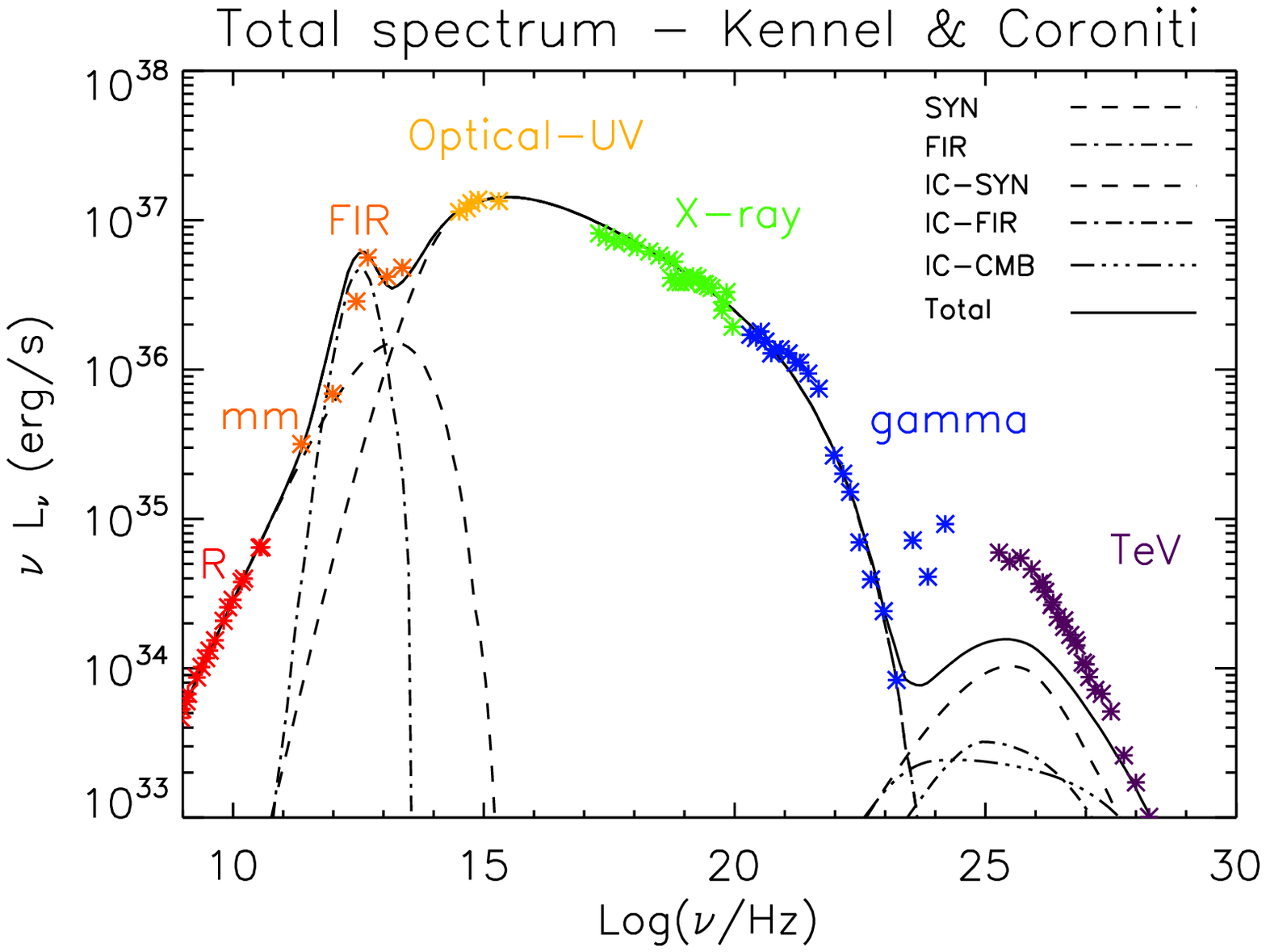}
}
\caption{Left panel: Kennel\&Coroniti total electron energy distribution function for different values of the maximum particle energy $\epsilon_{\infty}$ as specified in the figure. In the plot we use $E=\epsilon m c^{2}$ and $F=(f_{\mathrm{r}}+f_{\mathrm{w}})\cdot 4\mathrm{\pi}/(mc^{2})$ and the particle distribution function is normalized to an average value for the pressure of $2.1\times 10^{-9} \mathrm{erg~cm^{-3}}$. The dashed lines are the primordial and the wind component at $\epsilon_{\infty}\rightarrow\infty$. Right panel: Kennel\&Coroniti spectra are plotted against the data. The single contributions from different incident photon targets, thermal FIR emission and total luminosity are reported with different colors and line-styles. Dashed lines: primordial and wind synchrotron contributions and IC-SYN emission; dash-dotted line: thermal FIR and IC-FIR ones; dash-dot-dotted line: IC-CMB one; solid line: total luminosity. References to the observations reported here are given in the main text.  
}
\label{fig:kc_spectrum}
\end{figure*}
In order to understand the different contributions of the various physical processes to the overall emission of the Crab Nebula, simulated synchrotron and IC integrated emission spectra are obtained and plotted against the data from the radio to the gamma-ray band. 

From this comparison, we will derive the unknown parameters in the distribution functions defined in the previous section, within two different dynamical models: the steady-state, spherically symmetric model by \cite{kennel84a}, used as a test, and our axisymmetric 2-D simulations.

\subsection{The spherical model}
\label{sect:KC}
The emission recipes of Sect.~\ref{sect:model} are here applied to Kennel and Coroniti's spherical model, calculated for a wind magnetization parameter $\sigma=0.005$ \cite[as did ][]{atoyan96} and a nebular radius $R=2.0$\,pc. Integrated spectra in Fig.~\ref{fig:kc_spectrum} are compared with observations of the Crab Nebula and with the results by \cite{atoyan96}, used here as a benchmark for our model.

The values of the parameters correspond to best fit estimates (see Table~\ref{table: 1}). The primordial and wind spectral indices are equal to the observed values in the radio and optical bands \citep{baars72, veron93}, respectively. $A_{\mathrm{r}}$ and $A_{\mathrm{w}}$ are found by fitting the emission in the radio and X-ray band; $\epsilon^{*}_{\mathrm{r}}$ and $\bar{\epsilon}$ are chosen so as to connect the radio and optical parts of the spectrum smoothly; $\epsilon^{*}_{\mathrm{w}}$ is such as to reproduce the observations in the gamma-ray band. We recall here that $A_{\mathrm{0}}=A_{\mathrm{w}}p_0$ refers only to the non-thermal tail of the particle energy distribution. This accounts for some unknown fraction $\xi$ of the thermal pressure at the termination shock \citep{kennel84b}, with $\xi$ determined from:
\be
\label{eq:Aw}
\xi p_{0}=\frac{mc^{2}}{3}\int_{4\mathrm{\pi}}d\Omega \int_{\epsilon_{0}^{\mathrm{min}}}^{\epsilon_{0}^{\mathrm{max}}} f_{0}(\epsilon_{0})\epsilon_{0}\mathrm{d}\epsilon_{0}~~0<\xi<1,
\ee
where $\Omega$ is the solid angle, $\epsilon^{\mathrm{min}}_{0}=\bar{\epsilon}$, $\epsilon^{\mathrm{max}}_{0}=\epsilon^{*}_{\mathrm{w}}$ and $\mathrm{f}_{0}(\epsilon_{0})$ contains $A_{0}$ and hence $p_{0}$. We find $\xi \approx 70\%$.

The values of our parameters are slightly different from those in the paper by \cite{atoyan96}, from which the form of our distribution function is taken. The main difference is in the cut-off energy  $\epsilon^{*}_{\mathrm{w}}$ and is due to our approximated formula~(\ref{eq:eps0}), where adiabatic losses are neglected (these are considered instead by means of the factor $p$ in Eq.~(\ref{eq:func_distr_w_approx})). However the difference involves only the synchrotron spectrum at gamma-ray energies, with no consequences for the IC radiation flux, to which electrons near the cut-off contribute negligibly (see also Fig.~\ref{fig:video1} and the discussion in Sect.~\ref{sect:variability}). 
\begin{table}
\caption{Values of the parameters for the spherical model (KC) and for two different set up of our axisymmetric RMHD simulations, runA (1) and runA (2). $A_{r}$ is in unit $\mathrm{cm^{-3}}$ and $A_{w}$ is in unit $\mathrm{erg^{-1}}$.}
\label{table: 1}
\centering
\begin{tabular}{c c c c}
\hline \hline
Parameters & KC & RunA (1) & RunA (2)\\ 
\hline
$\alpha_{\mathrm{r}}$ & 0.26 & 0.26 & 0.26\\ 
$\alpha_{\mathrm{w}}$ & 0.70 & 0.70 & 0.85\\
$A_{\mathrm{r}}$ & $2.90\times 10^{-6}$ & $1.35\times 10^{-5}$ & $1.35\times10^{-5}$ \\ 
$A_{\mathrm{w}}$ & $3.2\times 10^{8}$ & $2.6\times10^{9}$ & $3.2\times10^{11}$\\
$\epsilon^{*}_{\mathrm{r}}$ & $2.0\times 10^{5}$ & $3.0\times10^{5}$ & $3.0\times10^{5}$\\ 
$\epsilon^{*}_{\mathrm{w}}$ & $1.4\times 10^{10}$ & $\infty$ & $\infty$\\
$\bar{\epsilon}$ & $4.3\times 10^{5}$ & $5.0\times 10^{5}$ & $8.0\times10^{5}$\\
\hline
\end{tabular}
\end{table}

In the left panel of Fig.~\ref{fig:kc_spectrum} we plot the total evolved distribution function, given by Eqs.~(\ref{eq:func_distr_r}) and (\ref{eq:func_distr_w_approx}), for different values of $\epsilon_{\infty}$. The particle distribution function is normalized through a value of the local pressure corresponding to the volume average in the simulation. Since the relic population is assumed to be spatially homogeneous, the radial dependence is only included in the wind particle distribution function through the local values of $p(r)$ and $\epsilon_{\infty}(r)$, both provided by the model \cite{kennel84a,kennel84b}. The wind population displays the expected behavior: both the local value of the thermal pressure, to which the peak of the distribution is proportional, and the cut-off energy, providing the exponential decay, are decreasing functions of $r$. Our results are basically coincident with those shown in Fig.~3 of \cite{atoyan96} in spite of a different visualization (see the respective captions).

The overall synthetic spectrum is then calculated using the above distribution function with the dynamics taken from Kennel\&Coroniti's model. In the right panel of Fig.~\ref{fig:kc_spectrum} the spectral luminosity is plotted and the individual synchrotron contributions of the two populations are also shown. The synchrotron spectrum from radio to IR frequencies is due to the primordial particles, homogeneously distributed all over the nebula. At higher frequencies the emission is due to the wind particles, which suffer from adiabatic and synchrotron cooling. The FIR thermal radiation is obtained from the black-body formula with a temperature of the emitting dust of $T=46\,\mathrm{K}$ \citep{strom92, green04}. As far as IC radiation is concerned, we find that the primordial electrons contribute to IC-CMB radiation up to $1.6\times 10^{22}$\,Hz, to IC-FIR up to $3.2\times10^{23}$\,Hz and to IC-SYN up to $7.9\times10^{23}$\,Hz \cite[in the paper by ][ the frequency value is larger by almost one order of magnitude]{atoyan96} mainly in the Thomson regime, $\mathrm{\epsilon}h\nu_{t} \ll mc^{2}$, while at higher frequencies the emission is due to scattering by the wind particles. 

References for the reported data are as follows: radio data are from \cite{baars72}; the mm data are from \cite{mezger86} and \cite{bandiera02}; the infrared points are from IRAS in the far to mid-infrared \citep{strom92} and from ISO in the mid to near infrared range \citep{douvion01}; optical is from \cite{veron93} and UV from \cite{hennessy92}. Points in the range between soft X and gamma-rays are taken from \cite{kuiper01}, who compiled data from BeppoSAX, COMPTEL and EGRET. In the TeV band we plot the data from MAGIC \citep{albert08} and HEGRA \citep{aharonian04}.

Shapes and values of the single curves, especially IC-SYN and IC-CMB, are consistent with those of Figs.~9 of \cite{atoyan96}, so we conclude that in our model the computation of inverse Compton emission seems reliable. We recall that the main simplifications introduced here are in the wind distribution function, Eq.~(\ref{eq:eps0}), and the homogeneous emissivity in the IC-SYN target photon density, Eq.~(\ref{eq:phot_dens_IC}). 

The computed synchrotron emission is in agreement with the data, whereas the IC luminosity is lower at all gamma-ray frequencies. In both the present 1-D calculation and in the model by \cite{atoyan96} the gamma-ray EGRET and TeV data can be simultaneously fitted only by invoking additional contributions to the emission namely Bremsstrahlung and hadronic processes. In addition one may tune the magnetic field strength, however this has only been done in an \emph{ad hoc} manner \citep{atoyan96,aharonian04} rather than changing the dynamics self-consistently.

The conclusion that the IC alone is insufficient to explain the gamma-ray fluxes from the Crab Nebula seems to be the common result of all 1-D models. However, 1-D stationary models are not able to account for a number of observed nebular properties, in particular the jet-torus morphology seen in the X-rays. Due to the oversimplified dynamics intrinsic to spherical models, the conclusions about the emission processes are also questionable. On the other hand, 2-D time-dependent RMHD models have proved to reproduce the inner structure of the Crab Nebula even in some of its finest details. In the following sections, we present a study of the synchrotron emission and, for the first time, of the IC gamma-ray radiation from PWNe in the context of the axisymmetric scenario.

\begin{figure*}
\centering
\resizebox{\hsize}{!}{
\includegraphics{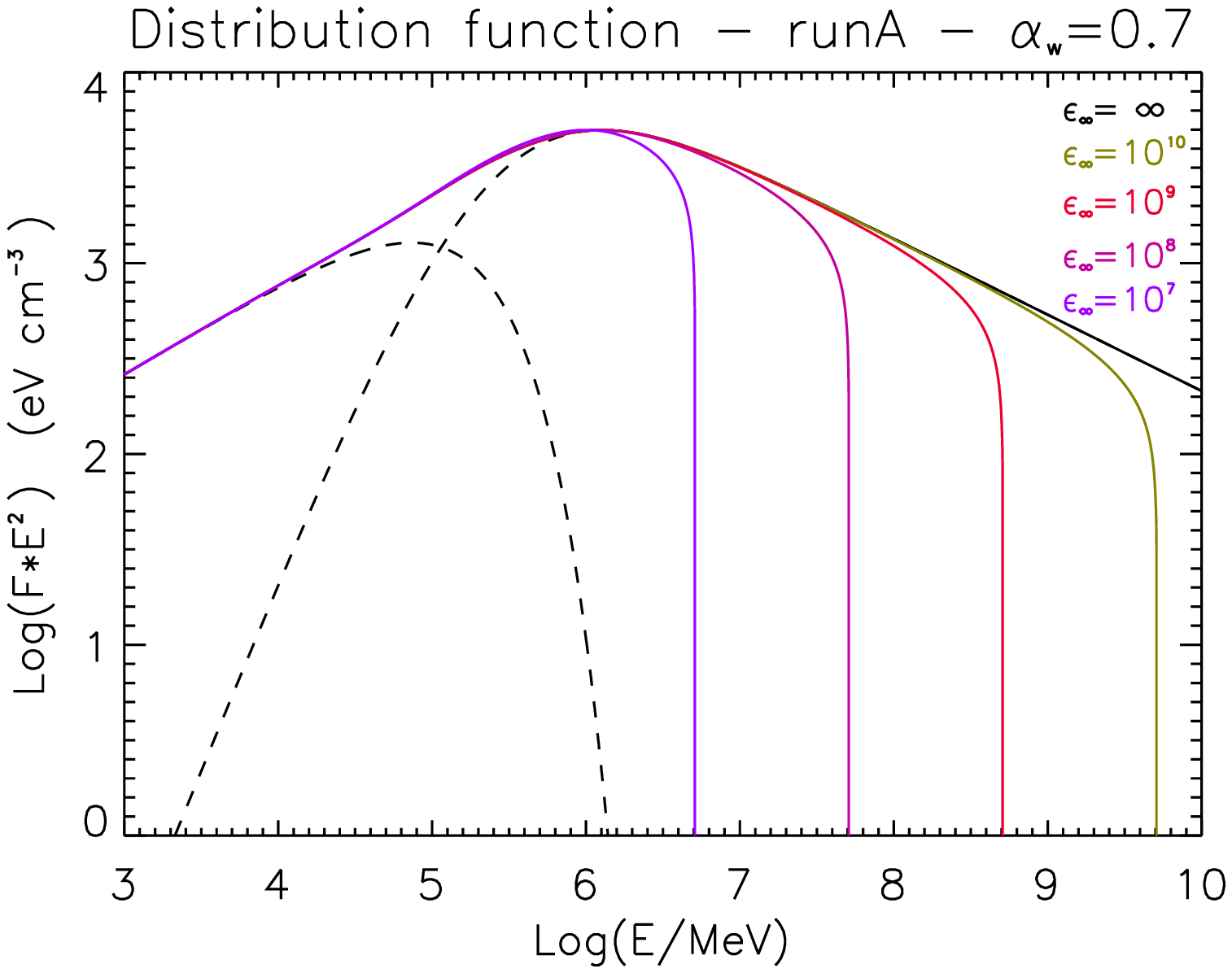}
 \includegraphics{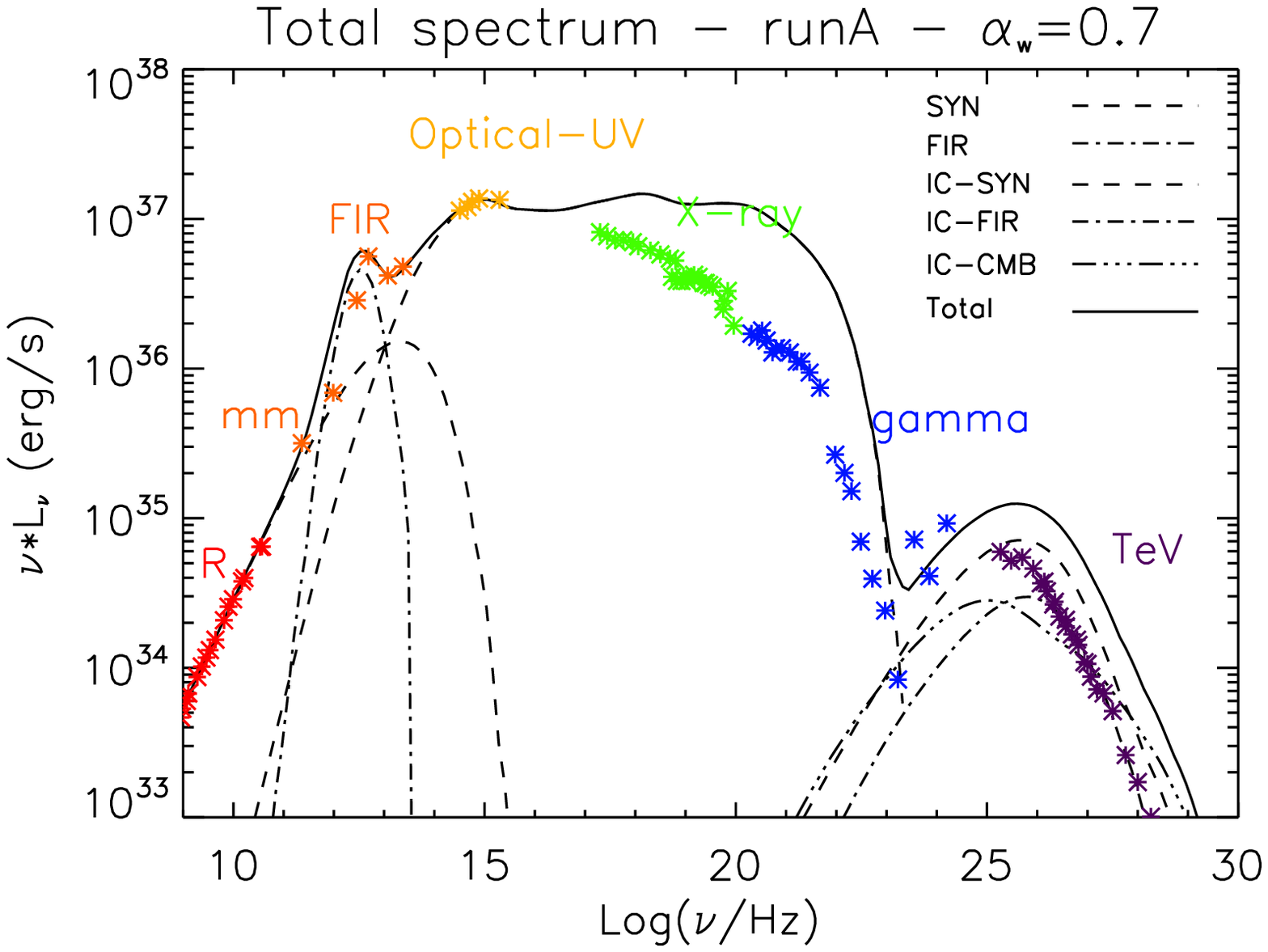}
}
\caption{Left panel: runA (1) total electron energy distribution function versus particle energy for different values of maximum particle energy $\epsilon_{\mathrm{\infty}}$ as specified in the figure. In the plot we use $E=\epsilon m c^{2}$ and $F=(f_{\mathrm{r}}+f_{\mathrm{w}})\cdot 4\mathrm{\pi}/(mc^{2})$ and the particle distribution function is normalized to an averaged pressure value of $2.1\times 10^{-9} \mathrm{erg~cm^{-3}}$. The dashed lines are the primordial and the wind component for $\epsilon_{\mathrm{\infty}}\rightarrow\infty$. Right panel: runA (1) spectra are plotted against the data. The single contributions from different incident photon targets, thermal FIR emission and total luminosity are reported with different colors and line-styles. Dashed lines: primordial and wind synchrotron contributions and IC-SYN emission; dash-dotted line: thermal FIR and IC-FIR ones; dash-dot-dotted line: IC-CMB one; solid line: total luminosity. References to the observations reported here are given in the main text.   
}
\label{fig:code_spectrum1}
\end{figure*}

\subsection{Results from numerical simulations}
\label{sect:sim}

\subsubsection{General discussion}

For the study of the non-thermal emission in the context of a 2-D model we focus here on the simulation labelled as runA in the paper by \cite{delzanna06}. The parameters that determine the shape and the strength of the magnetic field are: the anisotropy parameter, $\alpha=0.1$, the width of the \emph{striped} wind region, $b=10$ (corresponding to $\approx 10^{\circ}$ around the equator), and the wind magnetization, $\sigma=0.025$. The particles' maximum energy at injection, $\epsilon_{\infty}$, is here taken as $10^{10}$, corresponding to $5\times 10^{15}$\,eV \cite[see e.g. ][]{dejager92,aharonian04}.

\begin{figure*}
\centering
\resizebox{\hsize}{!}{
\includegraphics{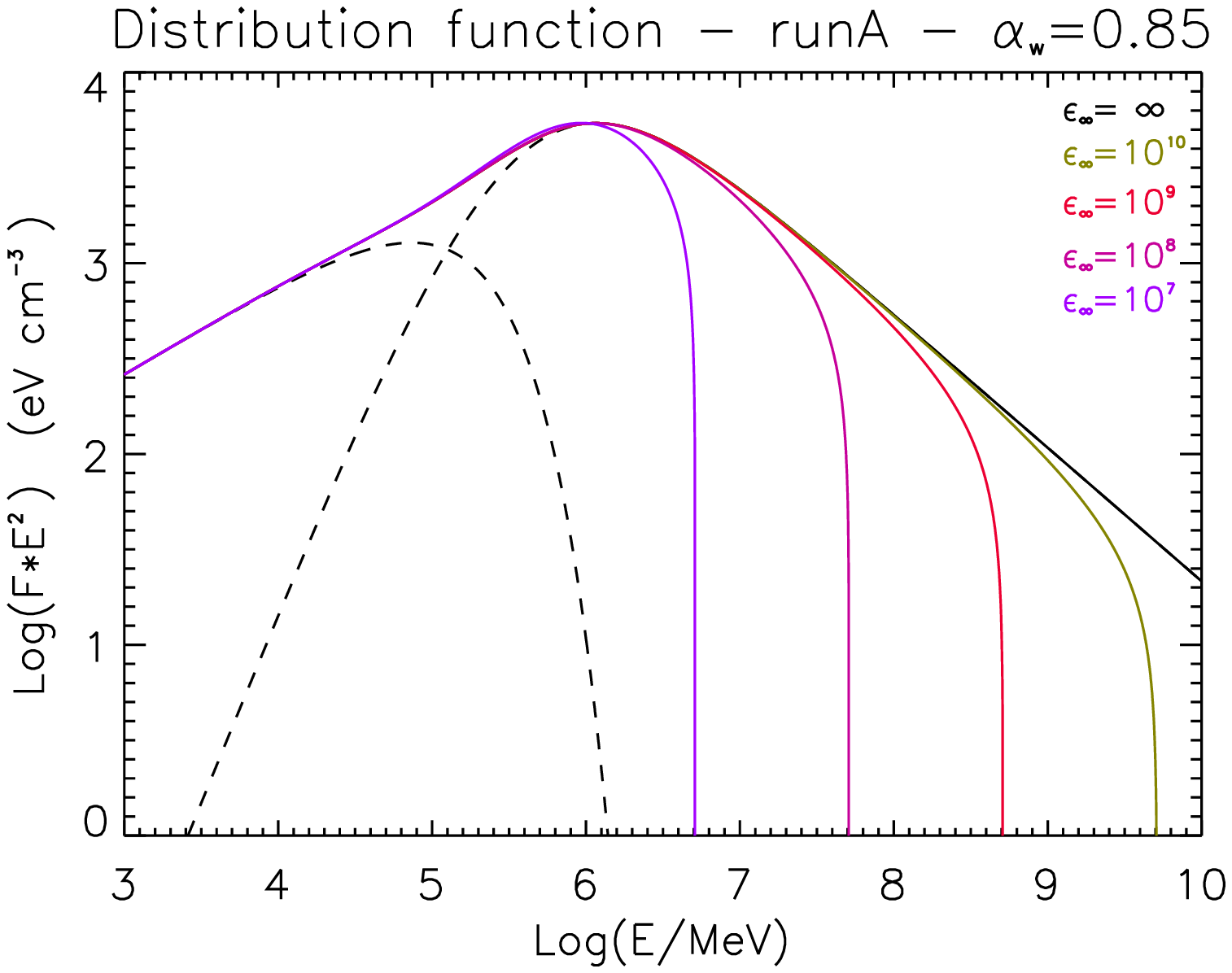}
 \includegraphics{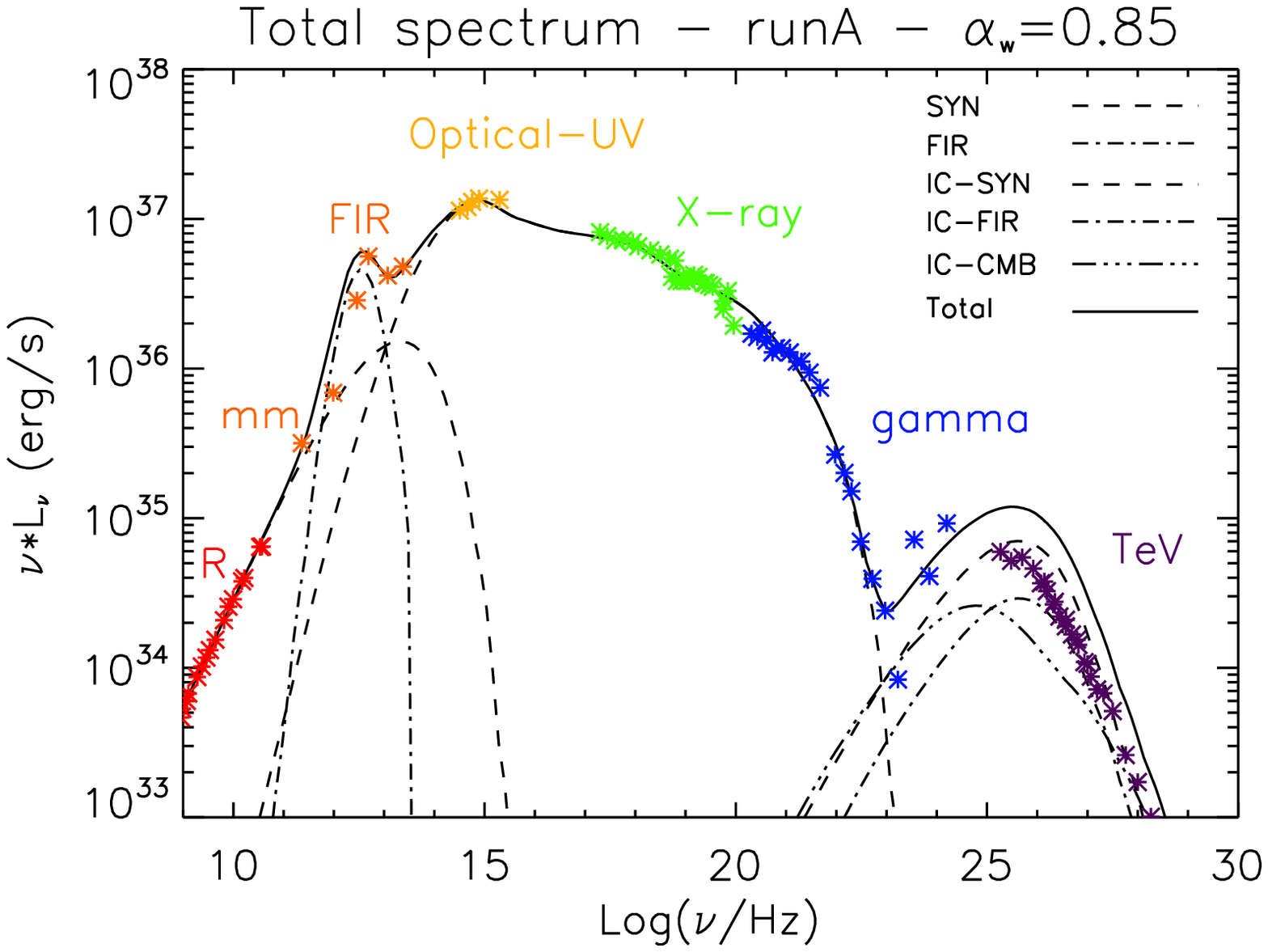}
}
\caption{The same as Fig.~\ref{fig:code_spectrum1} for runA (2).
}
\label{fig:code_spectrum2}
\end{figure*}

We now discuss the nebular emission spectra resulting from our simulation. Two different distribution functions are considered for the wind particles: one with the wind spectral index $\alpha_{\mathrm{w}}=0.7$, runA (1), and the other with $\alpha_{\mathrm{w}}=0.85$, runA (2). The first one is the counterpart of the spherical model computed in the previous section, and corresponds to the best fit optical spectral index \citep{veron93}. However, with this value of $\alpha_{\mathrm{w}}$ we are far from fitting the synchrotron spectrum from X to gamma-ray frequencies and the IC emission exceeds the data, as it is clear from the right panel of Fig.~\ref{fig:code_spectrum1}. A steeper injection spectrum is then adopted in order to discuss how the combined constraints from synthetic synchrotron and IC emission can help determining the physical parameters of PWNe. This second spectrum, in fact, allows one to reproduce the high energy synchrotron spectrum (see Fig.~\ref{fig:code_spectrum2}) but the TeV emission is slightly closer, though still above the data. In both runs the spectrum around $10$\,GeV seems approximately to fit the EGRET data, that, we recall, are affected by large error bars \cite[][]{kuiper01}. Hopefully GLAST observations will allow to constrain better the models in this crucial range of frequencies. At the moment, however, there is no room for additional contributions to the gamma-ray emission, such as Bremsstrahlung and hadronic processes. Note that these results are opposite to what found in 1-D (see Sect.~\ref{sect:KC}), where the model underpredicts the IC peak emission of a factor $\approx 5$.
 
\begin{figure*}
\centering
\resizebox{\hsize}{!}{
\includegraphics{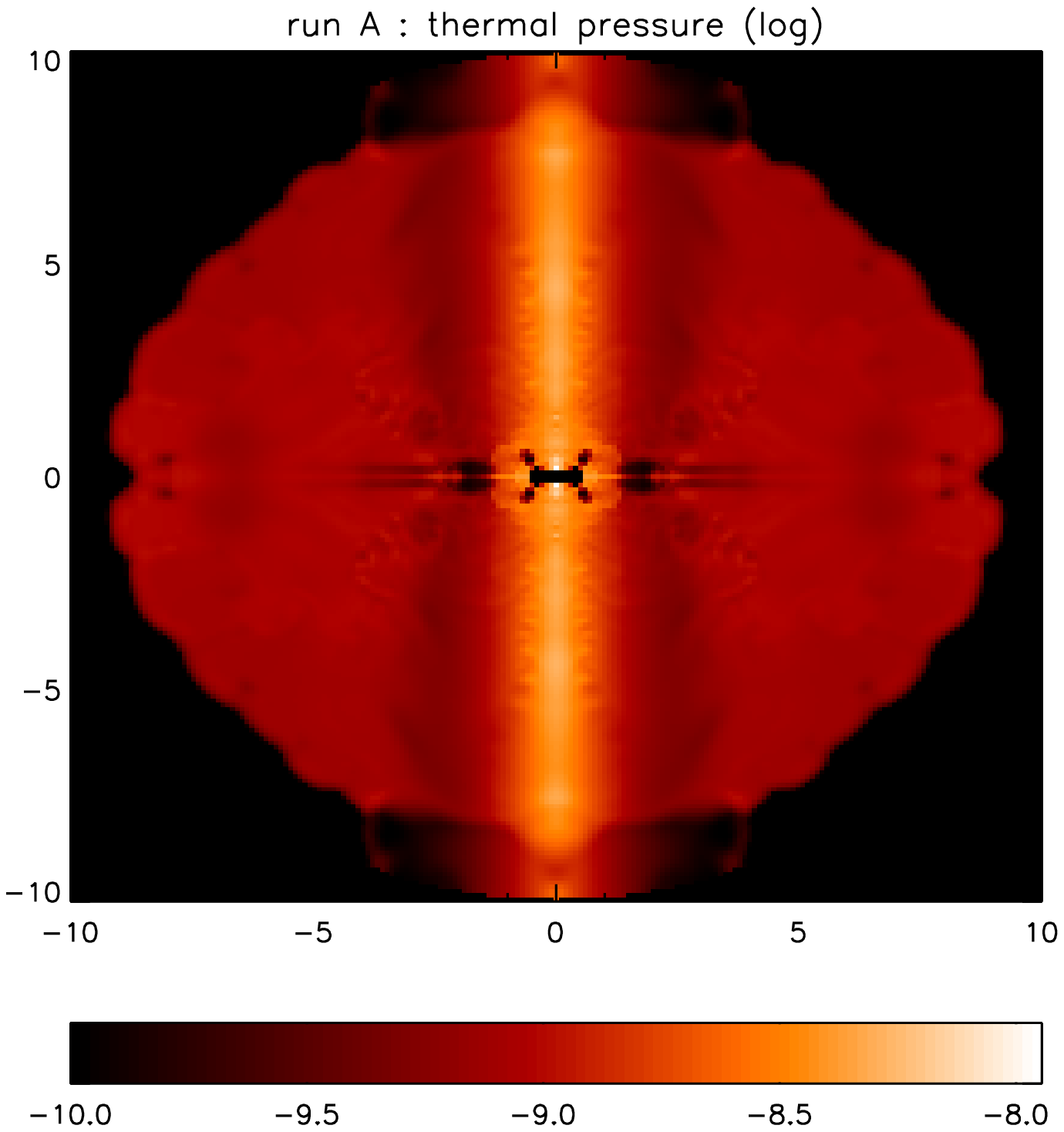}
\includegraphics{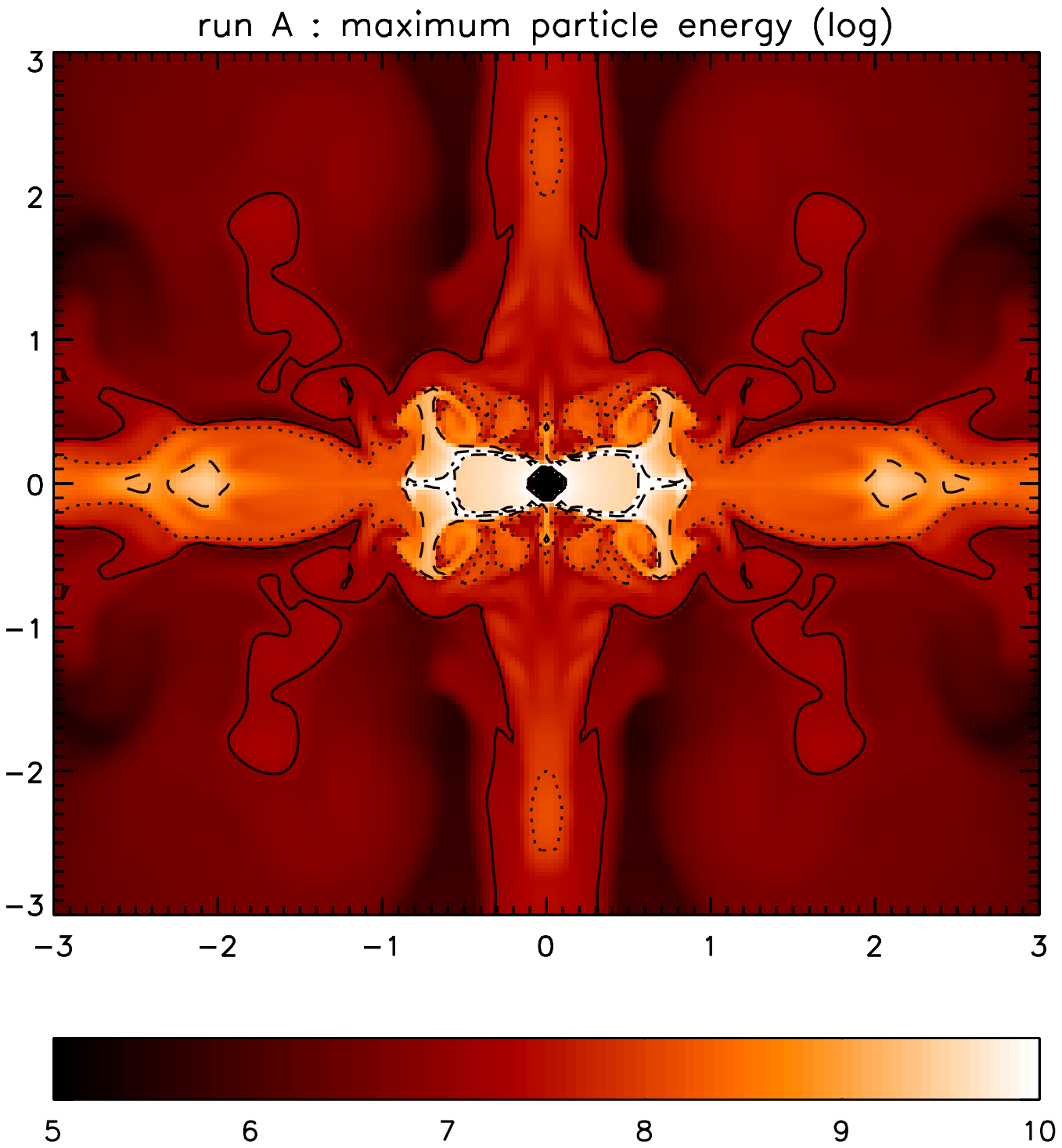}
}
\caption{Left panel: color coded map of the thermal pressure in units of $\mathrm{erg~cm}^{-3}$. Right panel: color coded map of the maximum particle energy; superimposed contours correspond to values of $10^{7}$ (solid line), $10^{8}$ (dotted line), $10^{9}$ (dashed line) and $10^{10}$ (dot-dashed line). Along the axes, distance from the central pulsar is reported in ly.   
}
\label{fig:pres_emax}
\end{figure*}

Before discussing the spectra in detail, let us first describe the particles distribution function. The total (primordial plus wind) particle distribution functions for the two cases, obtained as in Sect.~\ref{sect:electrons}, are shown in the left panels of Figs.~\ref{fig:code_spectrum1} and \ref{fig:code_spectrum2}. The values adopted for the maximum electron energy $\epsilon_{\infty}$ and for the thermal pressure are the same as those used in Fig.~\ref{fig:kc_spectrum}. However one should remember that these quantities are now functions of both the spherical radius $r$ and the polar angle $\theta$ (see Fig.~\ref{fig:pres_emax}). The remaining parameters are chosen such as to obtain with our simulated spectra the best fit possible of the observations (see Table~\ref{table: 1}).
 
By comparing the left panels of Figs.~\ref{fig:code_spectrum1} and \ref{fig:code_spectrum2} with the corresponding panel of Fig.~\ref{fig:kc_spectrum}, one can see that in our simulations a larger number of electrons is present at all energies. However, in runA (2) the distribution function falls down rapidly at high energies reaching values that are close to those found within the spherical model.

The values of the parameters adopted are reported in Table~\ref{table: 1}. Notice that for the assumed injection energy, $\epsilon_{\infty}$ is never larger than $10^{10}$. Hence the synchrotron cut-off in the gamma-rays is correctly reproduced without the need of introducing an exponential decay of the particle distribution function. 

For both cases, runA (1) and runA (2), the emission spectrum is calculated as described in Sect.~\ref{sect:model} and displayed in the right panels of Figs.~\ref{fig:code_spectrum1} and \ref{fig:code_spectrum2}. As far as the different contribution of the two populations in different frequency bands is concerned, the primordial electrons contribute to the synchrotron emission up to IR frequencies, to IC-CMB emission up to $2.5\times 10^{22}$\,Hz, to IC-FIR up to $3.2\times 10^{23}$\,Hz and to IC-SYN up to $7.9\times 10^{23}$\,Hz mainly in the Thomson regime. The radiation at higher frequencies for all components is due instead to the wind electrons. In both runs, the data are well fitted in the radio (by construction) and in the optical bands. In this context the first spectral break, between the IR and the optical, appears to be intrinsic and  due to the superposition of the two different populations of particles. The second one, in the UV band, corresponds in our model to a strong synchrotron burn-off and is related to the decreasing volume occupied by particles of increasing energy. Beyond $\approx 2\times 10^{15}$\,Hz, we observe a spectral flattening in our curves, followed by further minor breaks and counter-breaks at higher frequencies. Here, we recall that \cite{pavlov08} report X-ray observations of PWNe with spectral indices less than $0.5$, which have no theoretical explanation but that in principle could be reproduced by simulations similar to runA (1).

In the case of runA (1), at frequencies above $\approx 10^{16}$\, Hz the computed synchrotron emission is always far in excess of the data. The same happens at frequencies larger than $\approx 10^{24}$\, Hz, where inverse Compton is the dominant emission mechanism. Notice however that the IC-SYN component is not much affected by the hard X and gamma-ray synchrotron excess, which contributes only beyond $1$\,TeV. In runA (2) we manage to fit the X-ray data with our steeper simulated spectra, but still the IC emission appears to exceed the data by a factor $\sim 2$. 

These results show that the difficulties we have in reproducing the synchrotron spectrum with the commonly accepted value of $\alpha_{\mathrm{w}}$ cannot be solved by simply changing the particle distribution function, but are rather hinting to a problem related to the underlying dynamics, and in particular with the magnetic field distribution in the nebula.

In our simulations the nebular magnetic field appears to be compressed in localized regions close to the termination shock, while rapidly decreasing outward, much faster than in spherical models. Its volume averaged value is $B_{\mathrm{mean}}=1 \times 10^{-4}$\,G, which is $2-3$ times less than the value inferred from previous models \citep{kennel84a}. This means that in the outer regions, which occupy a large fraction of the volume and contribute proportionally to the integrated emission, our synthetic magnetic field is much lower than for spherical models. Hence, in order to fit the integrated spectra, we were forced to introduce a larger number of emitting electrons compared to the spherical model. Our electrons are also subject to less severe synchrotron losses on average, and consequently a steeper spectral index $\alpha_{\mathrm{w}}$, as in runA (2), is necessary in order to reproduce the high energy synchrotron spectrum. The simultaneous calculation of the IC spectrum, however, shows that the required large number of emitting electrons is not allowed by the gamma-ray data. Therefore, in order to reproduce the spectrum of the Crab Nebula, a larger and more distributed magnetic field would be needed, provided that it is able to reproduce the correct dynamics. Nevertheless before drawing any conclusions on the physics of the emission processes at work, a further investigation in the parameter space (for both the dynamics and the particle distributions) should be first attempted, since here we have only considered the runA results. Possibly the problem of the concentration of the nebular magnetic field is intrinsic to axisymmetric simulations and could be alleviated only by moving to 3-D.

\subsubsection{On the spectral slopes}

Let us now go back to the most surprising features observed in our spectra, namely the multiple changes of slope, which are in contrast with simple expectations from synchrotron cooling. However, the prediction of only one single steepening of the slope due to synchrotron burn-off is based on simplified 1-zone or 1-D nebular models, that assume a smooth flow and magnetic field distribution, so that particles experience continuous synchrotron losses. On the contrary, one of the main results of 2-D MHD models of PWNe has been the realization that the internal flow dynamics is more complex than previous 1-D models suggested.

This complexity reflects on the particle spectrum as a function of position within the nebula. As anticipated this is computed as described in Sect.~\ref{sect:electrons}, using the information on the thermal pressure $p$ and on the maximum energy $\epsilon_{\infty}$ provided by the simulations: color coded maps of $p$ and $\epsilon_{\infty}$ are shown in the left and right panel of Fig.~\ref{fig:pres_emax} respectively. 

The computed pressure tends to a constant with a spatial average of $2.1\times 10^{-9}\mathrm{erg~cm^{-3}}$ (which is behind our choice of the normalization constant in the plots showing the population distribution functions in Figs.~\ref{fig:kc_spectrum}, \ref{fig:code_spectrum1} and \ref{fig:code_spectrum2}). Exceptions are observed along the jets, where the maximum is located, and in the equatorial plane, where vortices create regions of lower pressure. 

As far as the local value of the maximum energy is concerned, the effects of synchrotron burn-off are highlighted by the four contours added on top of the color coded map of $\epsilon_{\infty}$. These correspond to the same values for which particle distribution functions are shown in the left panels of Figs.~\ref{fig:kc_spectrum}, \ref{fig:code_spectrum1} and \ref{fig:code_spectrum2}. The main difference between the results of our 2-D simulations and spherically symmetric models is immediately apparent from the contours in the right panel of Fig.~\ref{fig:pres_emax}: the spatial domain occupied by particles with energies up to a certain value is made of disconnected regions. While in 1-D the contours would be connected and encircle progressively smaller regions of space with increasing values of $\epsilon_{\infty}$, here the flow structure and magnetic field distribution is such that this is not the case anymore. In fact in our simulations fast flow channels in regions of relatively low magnetic field may take high energy particles far from the termination shock without suffering important synchrotron losses. In addition to this, the spatial distribution of the magnetic field arising from our model, which is compressed by the flux vortices around the TS and the polar axis \cite[see ][]{delzanna04,delzanna06}, causes a non monotonic decline of the volume occupied by high energy particles. This is at the origin of the multiple changes of slope in the synthetic emission spectrum. In fact, it is interesting to notice that 2-D simulations performed with a lower value of the magnetization parameter give origin to much less structured flow patterns and spectra showing just one change of slope, analogous to those found from spherical models. At the same time, however, the jets disappear. On the other hand, based on the possibility we find here of having more than one spectral break due to synchrotron losses, one may speculate that a larger value of the magnetization parameter, sufficient to move the multi-slopes to IR frequencies, may allow to explain the entire synchrotron spectrum of the Crab Nebula with just one population of particles. However additional particle acceleration may take place in regions other than the termination shock, in particular the termination spots of high speed flows in the equatorial plane and along the jets, and gives rise to the changes in slope.

\subsubsection{On the SYN/IC estimate for $B_{\mathrm{mean}}$}

The ratio between synchrotron and IC observed luminosity, when these data are available, is often used to evaluate the average magnetic field strength in PWNe \cite[e.g. ][ and references therein]{dejager96,djannati07}. However such an estimate is based on the assumption that the magnetic field is uniformly distributed (or slowly varying as in the KC model). On the other hand the recent numerical results of PWNe have clearly shown (as also confirmed in this paper, see the discussion in this section) that the distribution of the magnetic field and of the maximum energy of the emitting particles can change quite dramatically within the nebula. 

It is thus of interest, not just to observers but also to theorists, to understand if, in the case of 2-D PWNe models, the magnetic field inferred from the ratio SYN/IC provides a good approximation of the true value, or a biased one, at different frequencies.

In particular we compare the estimate of the magnetic field from the ratio SYN/IC, that are computed from the synthetic spectra, with the average value of the magnetic field $B_{\mathrm{mean}}$ equal to $1 \times 10^{-4}$\,G in our RMHD simulations, on which the same spectra are based. The obtained discrepancy is of order $10\%$ at synchrotron frequencies below $10^{23}$\,Hz, while beyond $10^{28}$\,Hz the discrepancy becomes larger than a factor of $2$. This is due to the fact that in the high frequency range the emission comes from wind particles confined to a small volume around the TS, where the magnetic field is highly structured and higher than the average value.

\section{Gamma-ray surface brightness maps}

\begin{figure*}
\includegraphics[height=6.8cm,width=6cm]{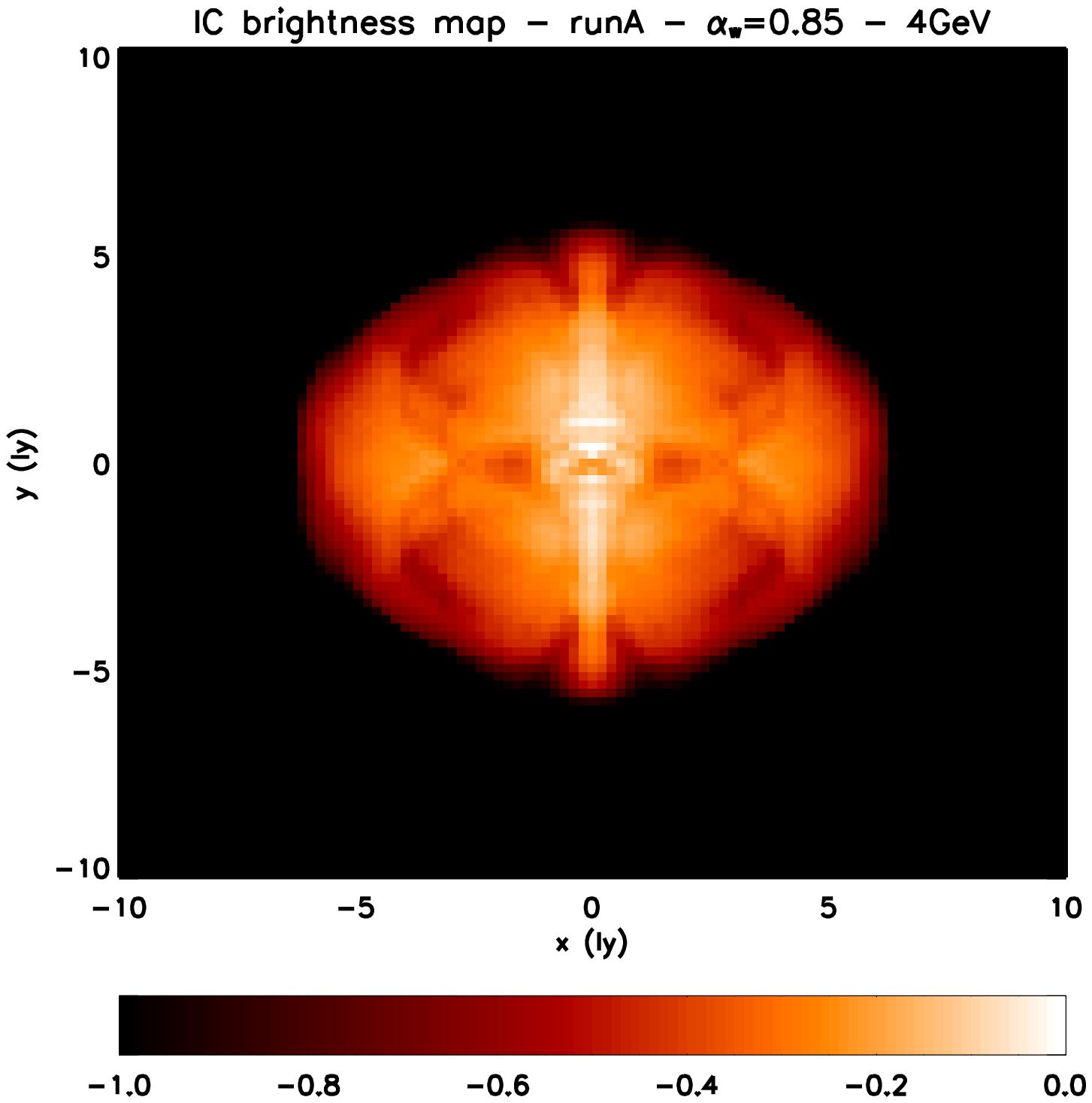}
\includegraphics[height=6.8cm,width=6cm]{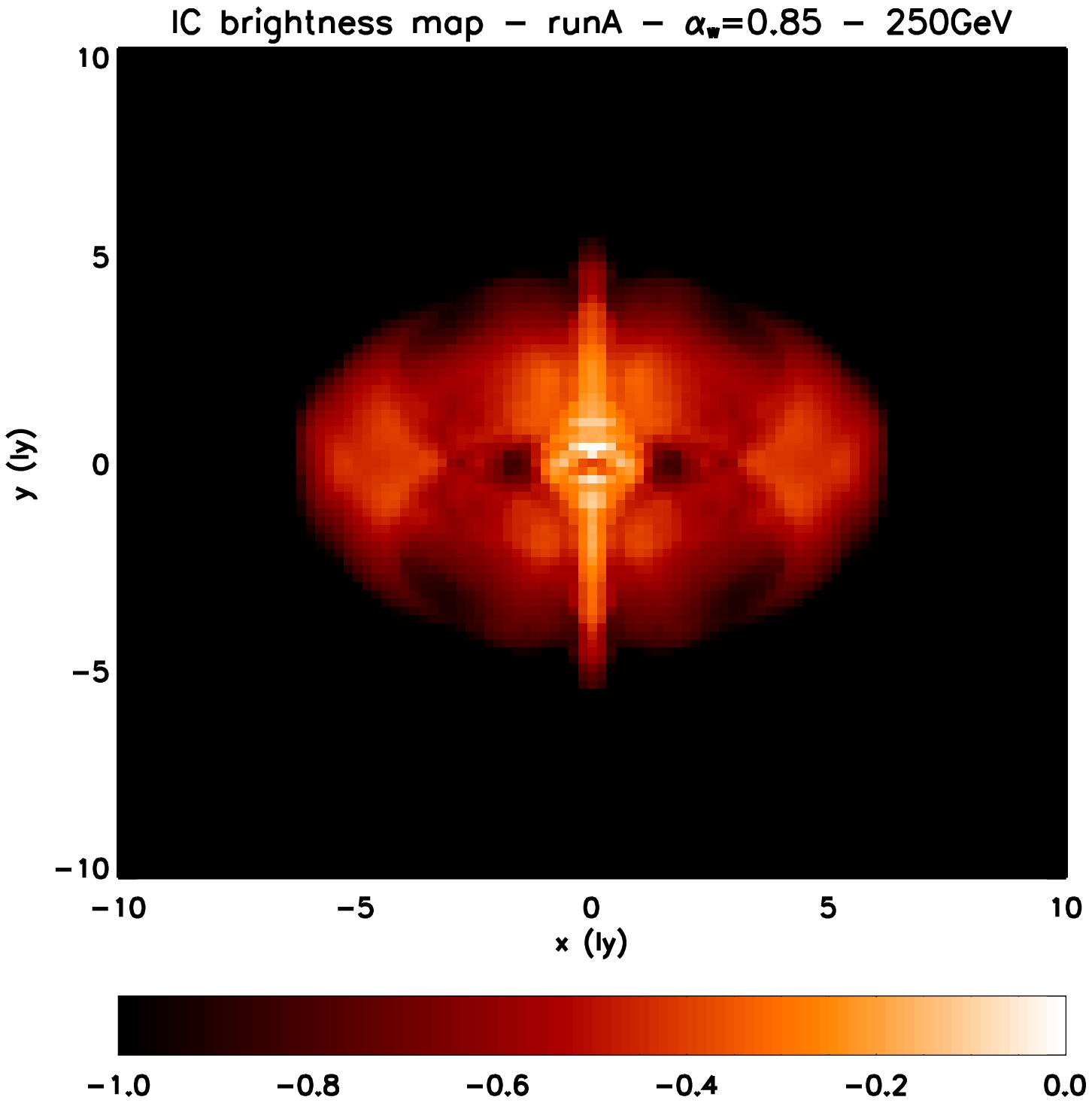}
\includegraphics[height=6.8cm,width=6cm]{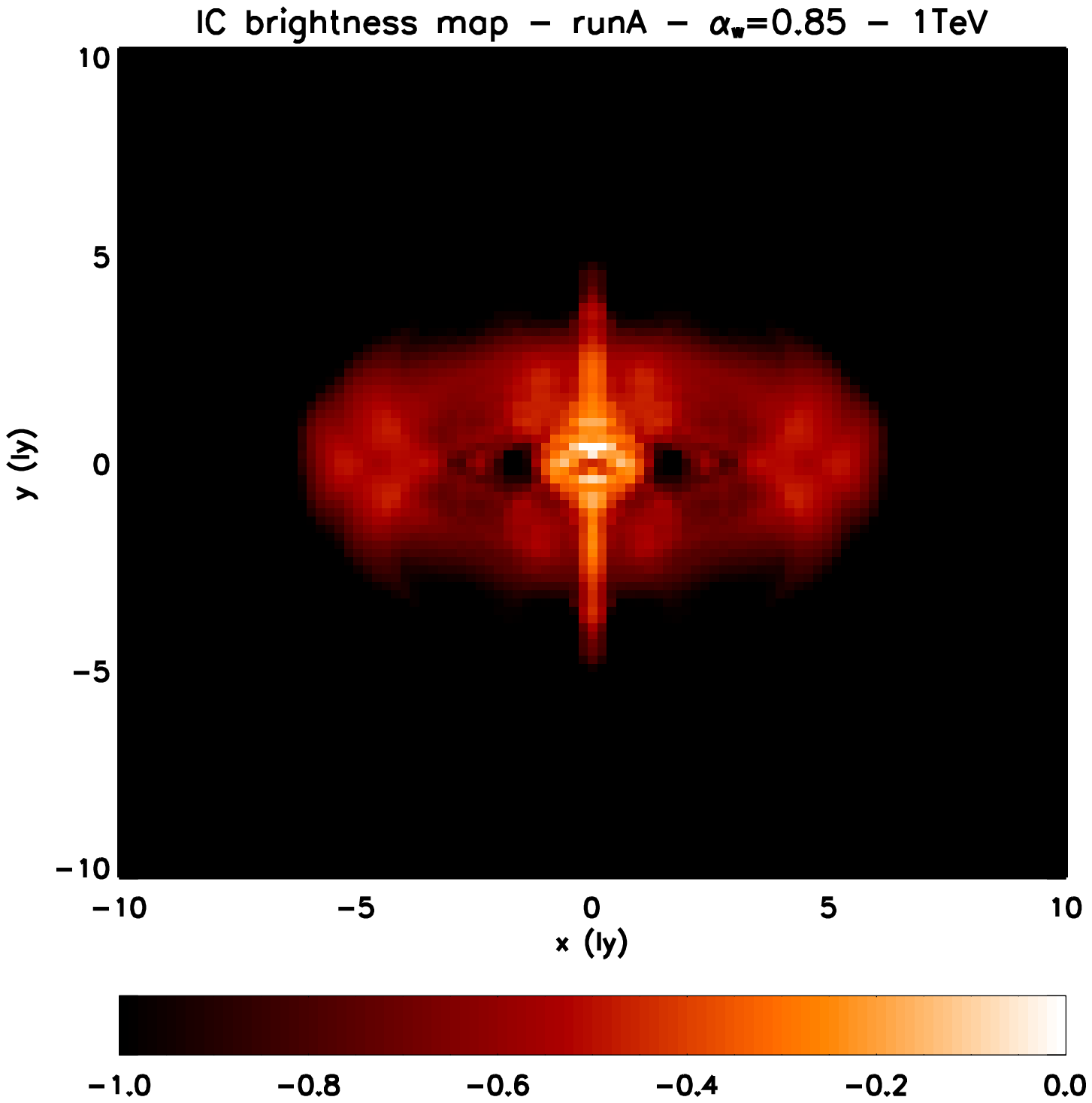}
\caption{RunA (2) case. Simulated total surface brightness maps in logarithmic scale normalized to the maximum at different IC frequency cut-offs: on the left at $4$\,GeV with a maximum $\approx 1.6\times 10^{-28}\mathrm{erg}~\mathrm{cm^{-2}~sr^{-1}~s^{-1}~Hz^{-1}}$; in the middle at $250$GeV with a maximum $\approx 1.0\times 10^{-29}\mathrm{erg}~\mathrm{cm^{-2}~sr^{-1}~s^{-1}~Hz^{-1}}$; on the right at $1$\,TeV with a maximum $\approx 1.8\times 10^{-30}\mathrm{erg}~\mathrm{cm^{-2}~sr^{-1}~s^{-1}~Hz^{-1}}$. On the axes: distance from the central pulsar in unit of ly.   
}
\label{fig:code_map}
\end{figure*}

\label{sect:maps}
In the present section, synthetic IC surface brightness maps are shown, for the first time, in the gamma-ray band. These are calculated for runA (2) by integrating along the line of sight the total (IC-SYN, IC-FIR and IC-CMB) emissivity, where the Doppler relativistic effects are included as explained in the paper by \cite{delzanna06}. 

In Fig.~\ref{fig:code_map} images at photon energies of $4$\,GeV, $250$\,GeV and $1$\,TeV are computed. These are chosen to lie in the middle of the band which will be observed by GLAST ($20$\,MeV-$300$\,GeV), in the range measured by MAGIC ($60$\,GeV-$9$\,TeV), and at the central energy of the HESS band ($100$\,GeV-$10$\,TeV). When comparing the three maps it is immediately apparent that the size of the nebula is larger at lower frequency.

The second feature is that the \emph{jet-torus} structure, common among the observed PWNe in the X-rays and here produced by the simulated RMHD evolution in time and space with a $\sigma_{\mathrm{eff}} \ga 0.01$, is clearly visible also at gamma-ray frequencies. One can distinguish the bright features as the central knot and the arcs, although the brightness contrast is now less than in the X-rays \cite[see ][]{delzanna06}. This similarity between the gamma-ray and X-ray morphology is due to the fact that the local emitting electrons are the same for both bands. 

An investigation of these bright features in our synthetic surface brightness maps at various synchrotron frequencies has shown that the central knot and the arcs near the TS appear at all energies and completely dominate the gamma-ray emission. Larger structures such as the jets and the torus start to fade away from the images at $\ga 1$\,keV and $\approx 10$\,keV, respectively. Moving up in frequency from the synchrotron peak the overall PWN size keeps on decreasing reaching a minimum at the cut-off SYN frequency ($0.5$GeV in our model), where basically only the electrons at the TS contribute. On the other hand, the original (radio band) dimensions are recovered when IC becomes the dominant emission process due to the ubiquity of the radio electrons responsible for emission at GeV frequencies. When the wind electrons begin to dominate the emission, synchrotron burn-off effects start to play a role and the volume of the nebula decreases again. So all structures, which disappear in the hard X-ray synchrotron emission, are again visible in gamma-ray images up to TeV photon energies. This is due to the fact that the IC emission structure does not directly depend on the complex magnetic field distribution but only on the spatial behavior of the electron distribution function, determined by the local value of $p$ and $\epsilon_{\infty}$.

Present day gamma-ray instruments have insufficient spatial resolution to distinguish single features in the internal structure, so that a comparison between the simulations and the data can only be based on the size of the nebula. However, indications of asymmetries in PWNe are already seen by HESS \citep{aharonian05b,aharonian06b}.

MAGIC has measured an average radius of about $4.5$\,ly at $250$\,GeV and of about $3$\,ly at $500$\,GeV \citep{albert08}. The simulated maps at the corresponding frequencies (we do not display here the $500$\,GeV image) have dimensions comparable to the observed values along the $y$-axis, though along the $x$-axis we find a negligible shrinking of the size with growing frequency. This is due to the combined effect of our relatively low magnetic field and the presence of fast flow channels in the equatorial region advecting particles outward fast enough for energy losses to be negligible. The average IC fluxes, observed by MAGIC, are about half of those obtained in our simulations ($\approx 4\times 10^{-10}\mathrm{erg~cm^{-2}~s^{-1}}$). This corresponds to the excess of the synthetic IC emission already discussed when treating the spectra of Sect.~\ref{sect:sim}.

\section{X and gamma-ray time variability}
\label{sect:variability}

\begin{figure*}
\centering
\includegraphics[width=12.5cm,bb=0 0 720 370,clip]{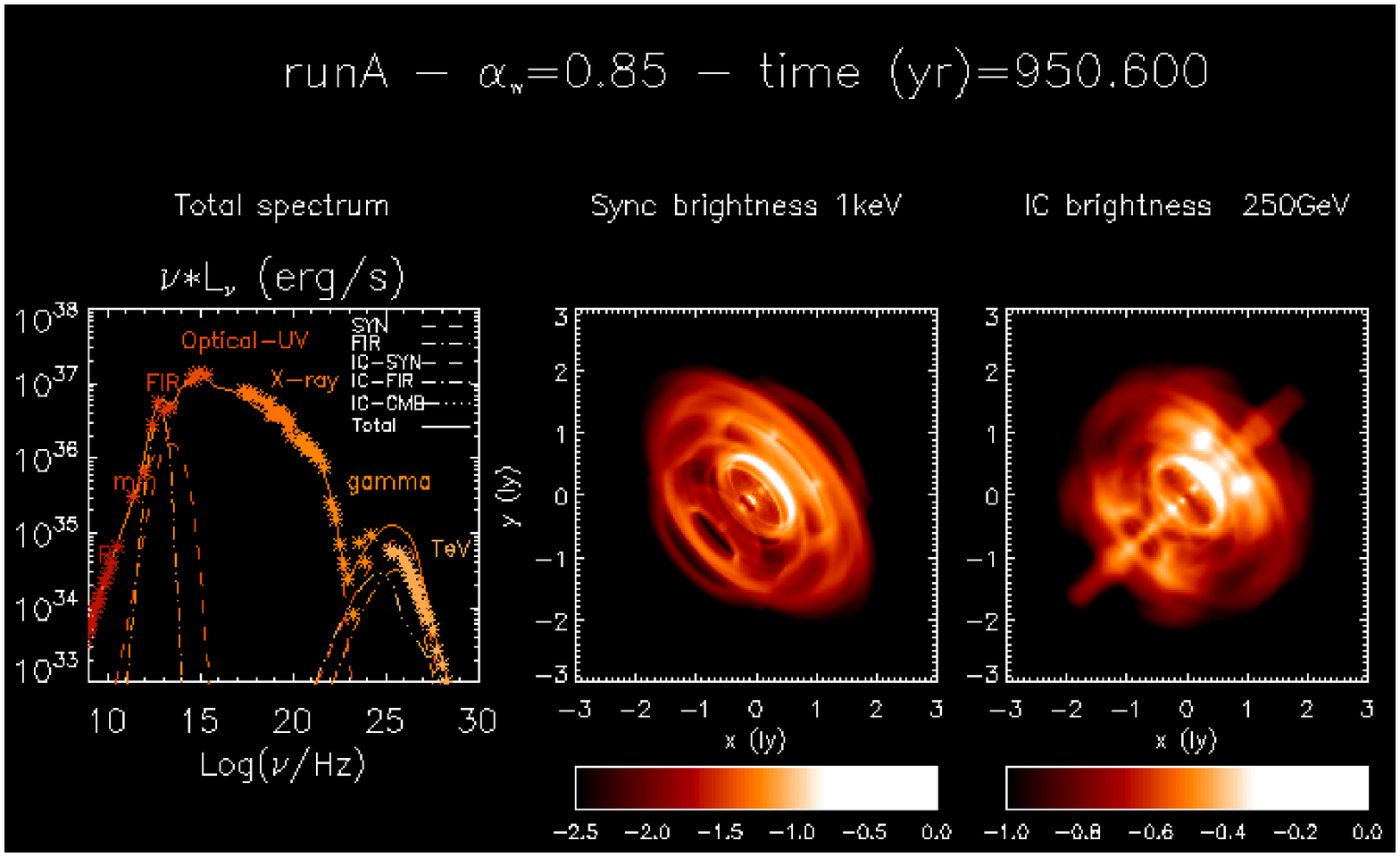}
\includegraphics[width=12.5cm,bb=0 0 720 370,clip]{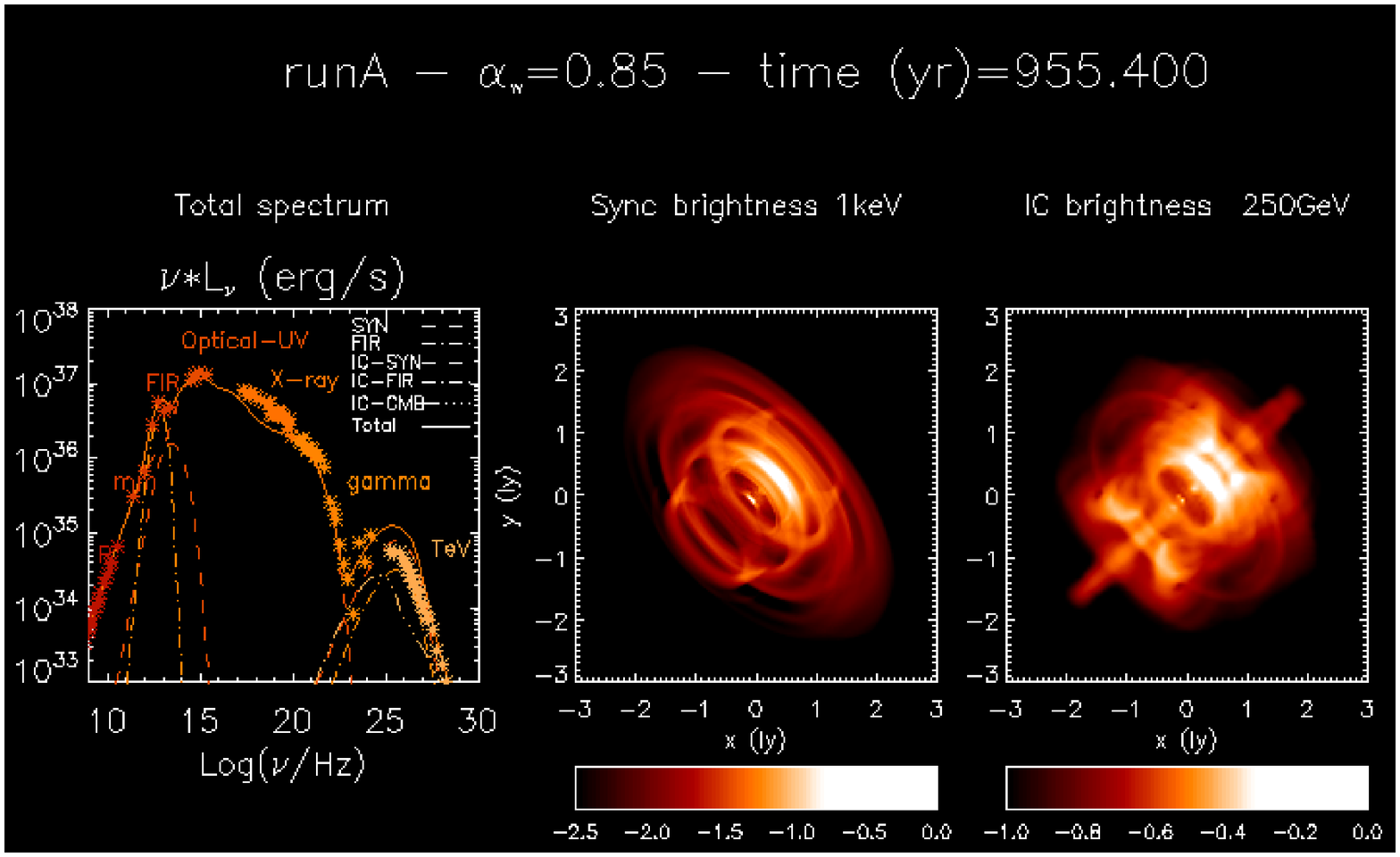}
\includegraphics[width=12.5cm,bb=0 0 720 370,clip]{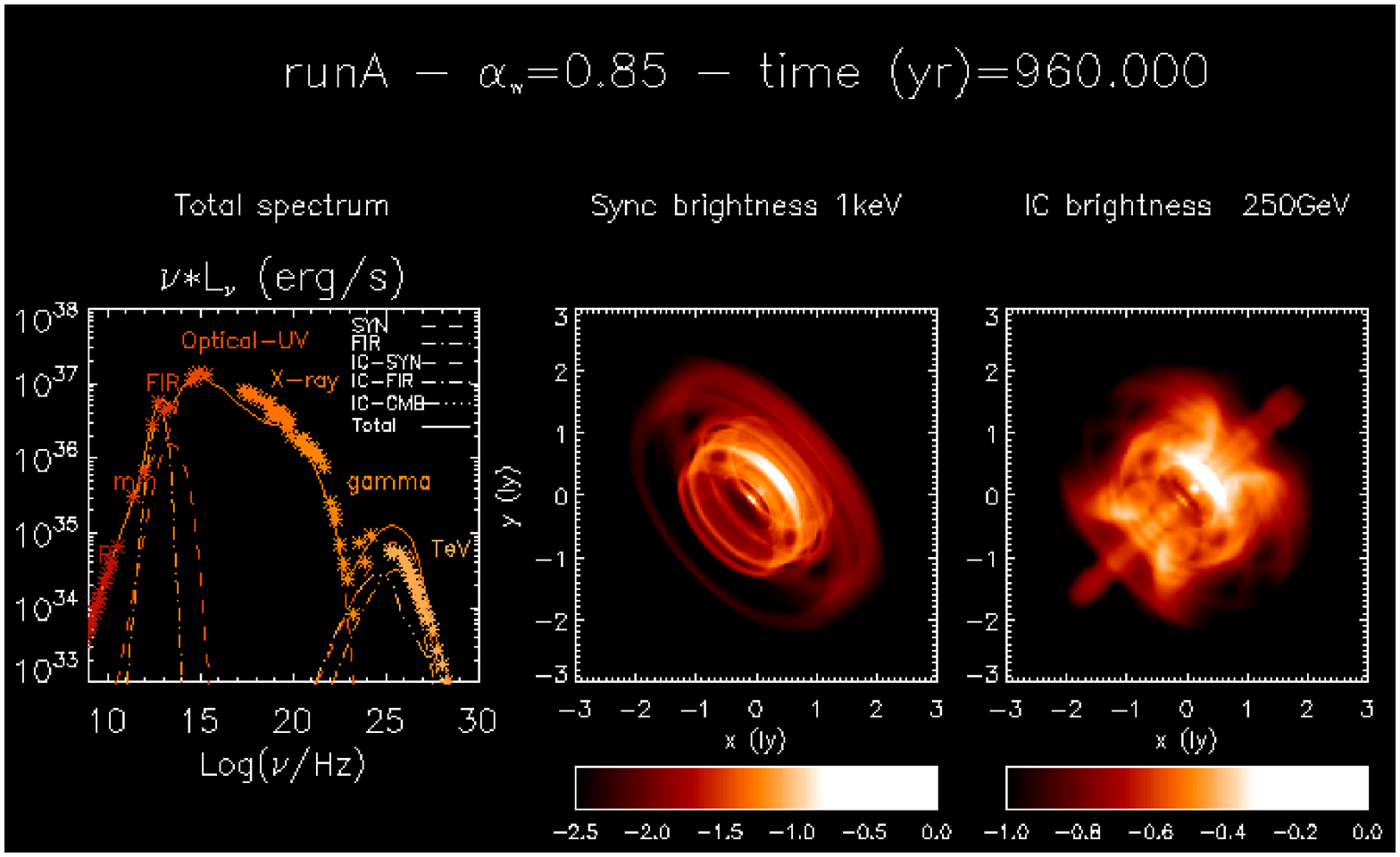}
\caption{RunA (2) case. Snapshots from a movie of PWN evolution: output times are selected at $950.6$\,yr, $955.4$\,yr and $960.0$\,yr. Results from one output time per row are shown. Left panels: plot of the synthetic spectral luminosity in logarithmic scale versus photon frequencies (log). Middle panels: simulated brightness maps (log) at $1$\,keV normalized to the maximum $6.2\times 10^{-18}\mathrm{erg}~\mathrm{cm^{-2}~sr^{-1}~s^{-1}~Hz^{-1}}$. Right panels: simulated brightness maps (log) at $250$\,GeV normalized to the maximum $1.2\times 10^{-29}\mathrm{erg}~\mathrm{cm^{-2}~sr^{-1}~s^{-1}~Hz^{-1}}$. The $x$ and $y$ axes of all surface brightness maps report the distance (in ly) from the central pulsar. Notice the inclination with respect to both the plane of the sky and the north direction (respectively $30^{\circ}$ and $48^{\circ}$ for the Crab Nebula) and that only the internal region (within a radius of $3$\,ly) is displayed.}
\label{fig:video1}
\end{figure*}

In a few PWNe, including the Crab Nebula, bright variable emission features are observed at small scales near the termination shock and in the torus. In the first case they are called \emph{wisps}, in the second filaments of the torus. The variability time-scale is of kilo-seconds in the IR band \cite[Gemini North Telescope, ][]{melatos05}, from days up to few months and from months up to a year, at optical and X-ray frequencies \cite[HST and Chandra, ][]{weisskopf00,hester02}, with time-scales that become longer with increasing distances from the TS. Quasi-periodically these structures are seen to appear, brighten very quickly and eventually fade, showing a characteristic outward motion. There are no corresponding gamma-ray observations because of the insufficient spatial resolution at those frequencies.

Different interpretations have been proposed in the literature for these time-variable features. A kinetic interpretation was given by \cite{spitkovsky04}, who showed that if protons are present in the pulsar wind, the variable structures can arise from ion-cyclotron instability at the shock front. Alternative explanations in the context of ideal MHD are based on nonlinear Kelvin-Helmholtz instabilities inside the nebula \citep{begelman99,bogovalov05,bucciantini06,bucciantini08}. Recently, \cite{bucciantini08a} presented simulations with \emph{wisps} near the TS moving outward at velocities of $\sim 0.5$\,c and slower filamentary structures in the torus, as observed. The period of the variability is of order one year, in agreement with Chandra X-ray observations. We are aware that Bucciantini and Komissarov are carrying out a detailed study of time variability within MHD models, in order to understand its origin, its characteristics and possible correlations with other nebular properties. Without going into details, we investigate here if the same time-dependent flow dynamics responsible for X-ray variability can lead to fluctuations in the gamma-ray emission \citep{dejager96,ling03} that might be of interest for future instruments (i.e. GLAST).

We consider runA (2) and compute snapshots of the integrated spectrum, synchrotron surface brightness map at $1$\,keV (Chandra band), IC surface brightness map at $250$\,GeV (in the middle of the gamma-ray band observed by MAGIC and towards the high energy limit of, but still within, the GLAST band). The sequence of images lasts ten years, from $950$\,yr to $960$\,yr, with intervals of 0.2\,yr. The full movie is downloadable from http://www.arcetri.astro.it/$\sim$delia/crab/runA2.gif, while a selection of snapshots is also shown in Fig.~\ref{fig:video1}.

Our simulations confirm the presence of variable \emph{wisps} near the TS and filamentary structures (rings) in the inner part of the torus, with characteristic time-scales of about $1$-$2$ years \cite[as observed for energies less than $0.75$\,MeV, ][]{much95}. These features move at speeds $0.3-0.5$\,c \cite[see also the flux velocity maps in ][]{delzanna06} and slow down before fading out at larger distances, in agreement with observations. This result suggests that variability in the inner nebula can be interpreted as associated to fluid motions occurring in the vicinities of the TS (either MHD compressive modes or Kelvin-Helmholtz instabilities, the latter arising at the shear flow boundaries). A part from these time-varying features, the most prominent structures, namely the central knot and the main arc, are instead stationary, also as observed.

The synthetic synchrotron surface brightness maps at X-ray frequencies and IC emission maps at $250$\,GeV photon energies show a strongly correlated variability, which is due to motion of the common parent electrons. However, as expected, since the IC emission is more uniformly distributed in the nebula than high energy synchrotron emission by the same electrons, the small-scale moving features are less evident compared to X-ray maps. The variability of the integrated IC spectrum is correspondingly reduced. We would like to recall that in our model at photon energies up to $1$\,TeV, the IC emission is found to be due to wind electrons up scattering radio-IR photons, therefore a study of the time-variations in this band, still makes sense in spite of the simplified assumptions underlying our computation of the IC emission. 
We would like to recall that we neglect all propagation effects in space and time, approximating the energy density of the target photon field as uniform over the nebular volume and slowly varying. Both these assumptions are satisfied in the radio-IR band.

\begin{figure}
\centering
\resizebox{\hsize}{!}{
\includegraphics{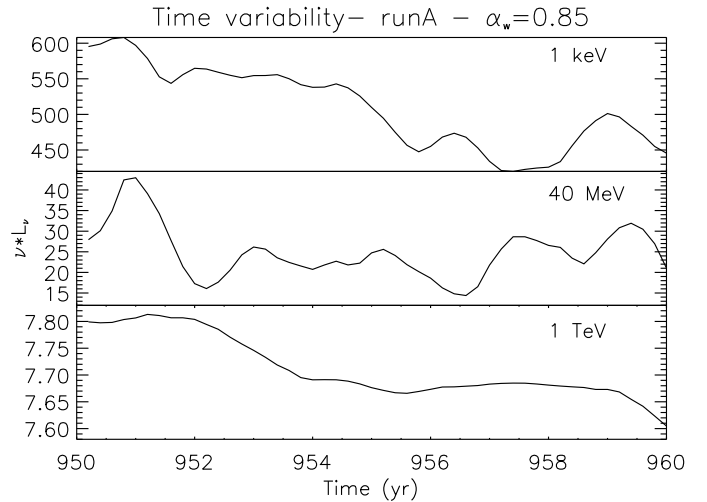}
}
\caption{RunA (2) case. Total luminosity $\nu L_\nu$ in unit $10^{34}$\,erg$\mathrm{s^{-1}}$s is displayed as a function of time, from $t$\,$=$\,$950$\,yr to $t$\,$=$\,$960$\,yr. From top to bottom panels, the corresponding photon energies are 1\,keV, 40\,MeV, 1\,TeV.   
}
\label{fig:glsattime}
\end{figure}

For a more quantitative analysis of the variability properties, in Fig.~\ref{fig:glsattime} we show the synthetic integrated emission as a function of time, for a selection of photon energies from X-rays to TeV gamma-rays ($1$\,keV, $40$\,MeV, $1$\,TeV). The strongest variations (about a factor of $2$) are seen to occur at synchrotron gamma-ray frequencies, where the emission is entirely due to the moving features close to the TS. On the other hand, the IC time series show very limited (about $1\%$) variations, as anticipated above. The results at synchrotron frequencies are in agreement with the observations by \cite{dejager96} who measured flux variations in the $1-150$\,MeV band, finding an amplitude of $30\%$ in the COMPTEL region ($1-30$\,MeV) and a factor of $2$ in the EGRET band ($70-150\,$MeV).

As far as time-scales are concerned, the gamma-ray variability, with a lower limit of order a few years in our simulations, also agrees with the results by \cite{dejager96}. However, the clear relationship between the moving features seen in the simulated surface brightness maps of Fig.~\ref{fig:video1} cannot be easily recovered in the luminosity time series, due to spatial integration effects. In any case, a combination of XMM/Chandra and GLAST observations might confirm the similarities between X and gamma-ray variability found in our simulations and consequently prove their common MHD origin.

\section{Conclusions}

In the present paper the work by \cite{delzanna06} on synchrotron emission from 2-D RMHD simulations of the PWNe (and of the Crab Nebula in particular) is extended to the gamma-ray frequencies, including the calculation of inverse Compton emission self-consistently. For the first time, maps of IC surface brightness are computed, to compare with existing (e.g. COMPTEL, HEGRA, HESS) and future observations (GLAST) of the Crab Nebula. Following recent results in the soft X-rays \citep{bucciantini08a}, a corresponding investigation of time variability in the gamma-ray band is also presented.

In order to compute the IC emission we include, following \cite{atoyan96}, a homogeneously distributed population of radio emitting electrons assumed born at the outburst of the SN in addition to the \emph{wind electrons}, continuously accelerated at the TS. In this scenario, the spectral break between IR and optical frequencies ($\approx 10^{13}$\,Hz) is intrinsic and due to the superposition of the two populations, while synchrotron burn-off is at the origin of the spectral break in the UV ($\approx 2\times 10^{15}$\,Hz).

From the comparison between our 2-D RMHD simulations and the Crab Nebula integrated spectra in Sect.~\ref{sect:sim}, we see that the value $\alpha_{w}=0.7$, suggested by the optical data of the inner parts of the nebula \citep{veron93} and adopted by \cite{atoyan96}, produces an overestimate of the synchrotron emission at frequencies above $10^{16}$\,Hz. Only with the spectral index $\alpha_{w}=0.85$ it is possible to reconcile the simulated emission with high energy synchrotron data of the Crab Nebula. However the computed IC emission exceeds the gamma-ray data by a factor of 2, showing that the steeper particle spectrum is not sufficient to remove the discrepancy between the results of the simulations and the data. The problem is related to the magnetic field in the simulations which is compressed towards the termination shock and lower on average than the estimates that are found in the literature, and likely than the value appropriate for the Crab Nebula. As a consequence, in order to reproduce the data we are forced to consider a larger number of particles. This seems to solve the problems as far as only the synchrotron emission is considered, but its inadequacy is then immediately revealed if calculation of IC is added. This suggests to adopt for the simulations a higher value of the pulsar wind magnetization with a word of caution that should be spent on the fact that is not clear to us how this would affect the nebular morphology, which is at present well reproduced. Another new result is that the complex nebular dynamics, highlighted by our 2-D simulations, gives rise to multi-slopes in the simulated synchrotron spectra, which are not expected in 1-D models.

Further insight can be gained from the gamma-ray morphology. Our synthetic maps show nebular dimensions and the shrinking of the size with increasing frequency in the range of those observed by MAGIC \citep{albert08} only along the polar axis, while dimensions do not change appreciably along the equator. This is due to the fast flow channels developing in a region of low magnetization: these advect the electrons far from the termination shock with negligible synchrotron losses. 

The simulated IC surface brightness maps are copies of the synchrotron X-ray ones, since they are produced by the same parent electrons. However, structures such as the jets and the external torus, that disappear in the hard X-rays, survive in the entire range of gamma-ray frequencies due to the fact that now the emission is not directly affected by the local magnetic field.

The X and gamma-ray variability of the inner nebula seems to have MHD origins. We find in our simulations characteristic time-scales of about $\approx 1-2$\,yr for both of them and time series of the integrated IC luminosity with smaller oscillations and slightly longer periods with respect to the synchrotron ones.

Future developments of the present work will consist firstly in treating more accurately the particle evolution in the nebula, for example by evolving directly the distribution function itself along the streamlines. The possibility of a dependence on the polar angle along the TS of the injected particle distribution function can be thus investigated. On the other hand, in order to overcome the problem of the magnetic field structure and strength, a wider investigation of the parameter space is necessary (we recall that our results pertain to a single simulation run). If this morphology will be confirmed, a problem intrinsic to the adopted axisymmetric geometry would be then highlighted and 3-D simulations should be attempted.

This analysis could also be extended to other PWNe than the Crab Nebula, even in different evolutionary stages (e.g. Vela). We believe that the powerful diagnostic techniques described here, which complete those presented in \cite{delzanna06}, are sufficiently versatile to be applied to other classes of non-thermal emitting sources.

\begin{acknowledgements}
The authors thank Rino Bandiera, for interesting discussions, and the anonymous referee for helpful suggestions. This work has been partly supported by the Italian Ministry of University and Research under PRIN MIUR 2006 (Pacini), by ASI-INAF 2007 (Bandiera). N. Bucciantini was supported by NASA through Hubble Fellowship grant HST-HF-01193.01-A, awarded by the Space Telescope Science Institute, which is operated by the Association of Universities for Research in Astronomy, Inc., for NASA, under contract NAS 5-26555.
\end{acknowledgements}

\bibliographystyle{aa}
\bibliography{pwn}

\begin{thebibliography}{65}
\expandafter\ifx\csname natexlab\endcsname\relax\def\natexlab#1{#1}\fi

\bibitem[{{Aharonian} {et~al.}(2005{\natexlab{a}}){Aharonian}, {Akhperjanian},
  {Aye}, {Bazer-Bachi}, {Beilicke}, {Benbow}, {Berge}, {Berghaus},
  {Bernl{\"o}hr}, {Boisson}, {Bolz}, {Borgmeier}, {Braun}, {Breitling},
  {Brown}, {Bussons Gordo}, {Chadwick}, {Chounet}, {Cornils}, {Costamante},
  {Degrange}, {Djannati-Ata{\"i}}, {O'C.~Drury}, {Dubus}, {Ergin}, {Espigat},
  {Feinstein}, {Fleury}, {Fontaine}, {Funk}, {Gallant}, {Giebels}, {Gillessen},
  {Goret}, {Hadjichristidis}, {Hauser}, {Heinzelmann}, {Henri}, {Hermann},
  {Hinton}, {Hofmann}, {Holleran}, {Horns}, {de Jager}, {Jung}, {Kh{\'e}lifi},
  {Komin}, {Konopelko}, {Latham}, {Le Gallou}, {Lemi{\`e}re}, {Lemoine},
  {Leroy}, {Lohse}, {Marcowith}, {Masterson}, {McComb}, {de Naurois}, {Nolan},
  {Noutsos}, {Orford}, {Osborne}, {Ouchrif}, {Panter}, {Pelletier}, {Pita},
  {P{\"u}hlhofer}, {Punch}, {Raubenheimer}, {Raue}, {Raux}, {Rayner},
  {Redondo}, {Reimer}, {Reimer}, {Ripken}, {Rob}, {Rolland}, {Rowell},
  {Sahakian}, {Saug{\'e}}, {Schlenker}, {Schlickeiser}, {Schuster}, {Schwanke},
  {Siewert}, {Sol}, {Steenkamp}, {Stegmann}, {Tavernet}, {Terrier},
  {Th{\'e}oret}, {Tluczykont}, {Vasileiadis}, {Venter}, {Vincent}, {Visser},
  {V{\"o}lk}, \& {Wagner}}]{aharonian05a}
{Aharonian}, F., {Akhperjanian}, A.~G., {Aye}, K.-M., {et~al.}
  2005{\natexlab{a}}, \aap, 432, L25

\bibitem[{{Aharonian} {et~al.}(2005{\natexlab{b}}){Aharonian}, {Akhperjanian},
  {Aye}, {Bazer-Bachi}, {Beilicke}, {Benbow}, {Berge}, {Berghaus},
  {Bernl{\"o}hr}, {Boisson}, {Bolz}, {Braun}, {Breitling}, {Brown}, {Bussons
  Gordo}, {Chadwick}, {Chounet}, {Cornils}, {Costamante}, {Degrange},
  {Djannati-Ata{\"i}}, {O'C.~Drury}, {Dubus}, {Emmanoulopoulos}, {Espigat},
  {Feinstein}, {Fleury}, {Fontaine}, {Fuchs}, {Funk}, {Gallant}, {Giebels},
  {Gillessen}, {Glicenstein}, {Goret}, {Hadjichristidis}, {Hauser},
  {Heinzelmann}, {Henri}, {Hermann}, {Hinton}, {Hofmann}, {Holleran}, {Horns},
  {de Jager}, {Kh{\'e}lifi}, {Komin}, {Konopelko}, {Latham}, {Le Gallou},
  {Lemi{\`e}re}, {Lemoine-Goumard}, {Leroy}, {Lohse}, {Martineau-Huynh},
  {Marcowith}, {Masterson}, {McComb}, {de Naurois}, {Nolan}, {Noutsos},
  {Orford}, {Osborne}, {Ouchrif}, {Panter}, {Pelletier}, {Pita},
  {P{\"u}hlhofer}, {Punch}, {Raubenheimer}, {Raue}, {Raux}, {Rayner},
  {Redondo}, {Reimer}, {Reimer}, {Ripken}, {Rob}, {Rolland}, {Rowell},
  {Sahakian}, {Saug{\'e}}, {Schlenker}, {Schlickeiser}, {Schuster}, {Schwanke},
  {Siewert}, {Sol}, {Steenkamp}, {Stegmann}, {Tavernet}, {Terrier},
  {Th{\'e}oret}, {Tluczykont}, {Vasileiadis}, {Venter}, {Vincent}, {V{\"o}lk},
  \& {Wagner}}]{aharonian05b}
{Aharonian}, F., {Akhperjanian}, A.~G., {Aye}, K.-M., {et~al.}
  2005{\natexlab{b}}, \aap, 435, L17

\bibitem[{{Aharonian} {et~al.}(2007){Aharonian}, {Akhperjanian}, {Bazer-Bachi},
  {Behera}, {Beilicke}, {Benbow}, {Berge}, {Bernl{\"o}hr}, {Boisson}, {Bolz},
  {Borrel}, {Braun}, {Brion}, {Brown}, {B{\"u}hler}, {B{\"u}sching},
  {Boutelier}, {Carrigan}, {Chadwick}, {Chounet}, {Coignet}, {Cornils},
  {Costamante}, {Degrange}, {Dickinson}, {Djannati-Ata{\"i}}, {Domainko},
  {Drury}, {Dubus}, {Egberts}, {Emmanoulopoulos}, {Espigat}, {Farnier},
  {Feinstein}, {Fiasson}, {F{\"o}rster}, {Fontaine}, {Funk}, {Funk},
  {F{\"u}{\ss}ling}, {Gallant}, {Giebels}, {Glicenstein}, {Gl{\"u}ck}, {Goret},
  {Hadjichristidis}, {Hauser}, {Hauser}, {Heinzelmann}, {Henri}, {Hermann},
  {Hinton}, {Hoffmann}, {Hofmann}, {Holleran}, {Hoppe}, {Horns},
  {Jacholkowska}, {de Jager}, {Kendziorra}, {Kerschhaggl}, {Kh{\'e}lifi},
  {Komin}, {Kosack}, {Lamanna}, {Latham}, {Le Gallou}, {Lemi{\`e}re},
  {Lemoine-Goumard}, {Lohse}, {Martin}, {Martineau-Huynh}, {Marcowith},
  {Masterson}, {Maurin}, {McComb}, {Moulin}, {de Naurois}, {Nedbal}, {Nolan},
  {Noutsos}, {Olive}, {Orford}, {Osborne}, {Panter}, {Pedaletti}, {Pelletier},
  {Petrucci}, {Pita}, {P{\"u}hlhofer}, {Punch}, {Ranchon}, {Raubenheimer},
  {Raue}, {Rayner}, {Ripken}, {Rob}, {Rolland}, {Rosier-Lees}, {Rowell},
  {Ruppel}, {Sahakian}, {Santangelo}, {Saug{\'e}}, {Schlenker}, {Schlickeiser},
  {Schr{\"o}der}, {Schwanke}, {Schwarzburg}, {Schwemmer}, {Shalchi}, {Sol},
  {Spangler}, {Steenkamp}, {Stegmann}, {Superina}, {Tam}, {Tavernet},
  {Terrier}, {Tluczykont}, {van Eldik}, {Vasileiadis}, {Venter}, {Vialle},
  {Vincent}, {V{\"o}lk}, {Wagner}, \& {Ward}}]{aharonian07a}
{Aharonian}, F., {Akhperjanian}, A.~G., {Bazer-Bachi}, A.~R., {et~al.} 2007,
  \aap, 472, 489

\bibitem[{{Aharonian} {et~al.}(2006{\natexlab{a}}){Aharonian}, {Akhperjanian},
  {Bazer-Bachi}, {Beilicke}, {Benbow}, {Berge}, {Bernl{\"o}hr}, {Boisson},
  {Bolz}, {Borrel}, {Braun}, {Breitling}, {Brown}, {B{\"u}hler},
  {B{\"u}sching}, {Carrigan}, {Chadwick}, {Chounet}, {Cornils}, {Costamante},
  {Degrange}, {Dickinson}, {Djannati-Ata{\"i}}, {O'C.~Drury}, {Dubus},
  {Egberts}, {Emmanoulopoulos}, {Epinat}, {Espigat}, {Feinstein}, {Ferrero},
  {Fontaine}, {Funk}, {Funk}, {Gallant}, {Giebels}, {Glicenstein}, {Goret},
  {Hadjichristidis}, {Hauser}, {Hauser}, {Heinzelmann}, {Henri}, {Hermann},
  {Hinton}, {Hofmann}, {Holleran}, {Horns}, {Jacholkowska}, {de Jager},
  {Kh{\'e}lifi}, {Komin}, {Konopelko}, {Latham}, {Le Gallou}, {Lemi{\`e}re},
  {Lemoine-Goumard}, {Lohse}, {Martin}, {Martineau-Huynh}, {Marcowith},
  {Masterson}, {McComb}, {de Naurois}, {Nedbal}, {Nolan}, {Noutsos}, {Orford},
  {Osborne}, {Ouchrif}, {Panter}, {Pelletier}, {Pita}, {P{\"u}hlhofer},
  {Punch}, {Raubenheimer}, {Raue}, {Rayner}, {Reimer}, {Reimer}, {Ripken},
  {Rob}, {Rolland}, {Rowell}, {Sahakian}, {Saug{\'e}}, {Schlenker},
  {Schlickeiser}, {Schwanke}, {Sol}, {Spangler}, {Spanier}, {Steenkamp},
  {Stegmann}, {Superina}, {Tavernet}, {Terrier}, {Th{\'e}oret}, {Tluczykont},
  {van Eldik}, {Vasileiadis}, {Venter}, {Vincent}, {V{\"o}lk}, {Wagner}, \&
  {Ward}}]{aharonian06b}
{Aharonian}, F., {Akhperjanian}, A.~G., {Bazer-Bachi}, A.~R., {et~al.}
  2006{\natexlab{a}}, \aap, 448, L43

\bibitem[{{Aharonian} {et~al.}(2006{\natexlab{b}}){Aharonian}, {Akhperjanian},
  {Bazer-Bachi}, {Beilicke}, {Benbow}, {Berge}, {Bernl{\"o}hr}, {Boisson},
  {Bolz}, {Borrel}, {Braun}, {Brown}, {B{\"u}hler}, {B{\"u}sching}, {Carrigan},
  {Chadwick}, {Chounet}, {Cornils}, {Costamante}, {Degrange}, {Dickinson},
  {Djannati-Ata{\"i}}, {O'C.~Drury}, {Dubus}, {Egberts}, {Emmanoulopoulos},
  {Espigat}, {Feinstein}, {Ferrero}, {Fiasson}, {Fontaine}, {Funk}, {Funk},
  {F{\"u}{\ss}ling}, {Gallant}, {Giebels}, {Glicenstein}, {Goret},
  {Hadjichristidis}, {Hauser}, {Hauser}, {Heinzelmann}, {Henri}, {Hermann},
  {Hinton}, {Hoffmann}, {Hofmann}, {Holleran}, {Horns}, {Jacholkowska}, {de
  Jager}, {Kendziorra}, {Kh{\'e}lifi}, {Komin}, {Konopelko}, {Kosack},
  {Latham}, {Le Gallou}, {Lemi{\`e}re}, {Lemoine-Goumard}, {Lohse}, {Martin},
  {Martineau-Huynh}, {Marcowith}, {Masterson}, {Maurin}, {McComb}, {de
  Naurois}, {Nedbal}, {Nolan}, {Noutsos}, {Orford}, {Osborne}, {Ouchrif},
  {Panter}, {Pelletier}, {Pita}, {P{\"u}hlhofer}, {Punch}, {Raubenheimer},
  {Raue}, {Rayner}, {Reimer}, {Reimer}, {Ripken}, {Rob}, {Rolland}, {Rowell},
  {Sahakian}, {Santangelo}, {Saug{\'e}}, {Schlenker}, {Schlickeiser},
  {Schr{\"o}der}, {Schwanke}, {Schwarzburg}, {Shalchi}, {Sol}, {Spangler},
  {Spanier}, {Steenkamp}, {Stegmann}, {Superina}, {Tavernet}, {Terrier},
  {Th{\'e}oret}, {Tluczykont}, {van Eldik}, {Vasileiadis}, {Venter}, {Vincent},
  {V{\"o}lk}, {Wagner}, \& {Ward}}]{aharonian06c}
{Aharonian}, F., {Akhperjanian}, A.~G., {Bazer-Bachi}, A.~R., {et~al.}
  2006{\natexlab{b}}, \aap, 456, 245

\bibitem[{{Aharonian} {et~al.}(2006{\natexlab{c}}){Aharonian}, {Akhperjanian},
  {Bazer-Bachi}, {Beilicke}, {Benbow}, {Berge}, {Bernl{\"o}hr}, {Boisson},
  {Bolz}, {Borrel}, {Braun}, {Brown}, {B{\"u}hler}, {B{\"u}sching}, {Carrigan},
  {Chadwick}, {Chounet}, {Cornils}, {Costamante}, {Degrange}, {Dickinson},
  {Djannati-Ata{\"i}}, {O'C.~Drury}, {Dubus}, {Egberts}, {Emmanoulopoulos},
  {Espigat}, {Feinstein}, {Ferrero}, {Fiasson}, {Fontaine}, {Funk}, {Funk},
  {F{\"u}{\ss}ling}, {Gallant}, {Giebels}, {Glicenstein}, {Goret},
  {Hadjichristidis}, {Hauser}, {Hauser}, {Heinzelmann}, {Henri}, {Hermann},
  {Hinton}, {Hoffmann}, {Hofmann}, {Holleran}, {Horns}, {Jacholkowska}, {de
  Jager}, {Kendziorra}, {Kh{\'e}lifi}, {Komin}, {Konopelko}, {Kosack},
  {Latham}, {Le Gallou}, {Lemi{\`e}re}, {Lemoine-Goumard}, {Lohse}, {Martin},
  {Martineau-Huynh}, {Marcowith}, {Masterson}, {Maurin}, {McComb}, {Moulin},
  {de Naurois}, {Nedbal}, {Nolan}, {Noutsos}, {Orford}, {Osborne}, {Ouchrif},
  {Panter}, {Pelletier}, {Pita}, {P{\"u}hlhofer}, {Punch}, {Raubenheimer},
  {Raue}, {Rayner}, {Reimer}, {Reimer}, {Ripken}, {Rob}, {Rolland}, {Rowell},
  {Sahakian}, {Santangelo}, {Saug{\'e}}, {Schlenker}, {Schlickeiser},
  {Schr{\"o}der}, {Schwanke}, {Schwarzburg}, {Shalchi}, {Sol}, {Spangler},
  {Spanier}, {Steenkamp}, {Stegmann}, {Superina}, {Tavernet}, {Terrier},
  {Th{\'e}oret}, {Tluczykont}, {van Eldik}, {Vasileiadis}, {Venter}, {Vincent},
  {V{\"o}lk}, {Wagner}, \& {Ward}}]{aharonian06d}
{Aharonian}, F., {Akhperjanian}, A.~G., {Bazer-Bachi}, A.~R., {et~al.}
  2006{\natexlab{c}}, \aap, 460, 365

\bibitem[{{Aharonian}(2007)}]{aharonian07}
{Aharonian}, F.~A. 2007, Science, 315, 70

\bibitem[{{Aharonian} {et~al.}(2004){Aharonian}, {Akhperjanian}, {Beilicke},
  {Bernl{\"o}hr}, {B{\"o}rst}, {Bojahr}, {Bolz}, {Coarasa}, {Contreras},
  {Cortina}, {Denninghoff}, {Fonseca}, {Girma}, {G{\"o}tting}, {Heinzelmann},
  {Hermann}, {Heusler}, {Hofmann}, {Horns}, {Jung}, {Kankanyan}, {Kestel},
  {Kohnle}, {Konopelko}, {Kranich}, {Lampeitl}, {Lopez}, {Lorenz}, {Lucarelli},
  {Mang}, {Mazin}, {Meyer}, {Mirzoyan}, {Moralejo}, {O{\~n}a-Wilhelmi},
  {Panter}, {Plyasheshnikov}, {P{\"u}hlhofer}, {de los Reyes}, {Rhode},
  {Ripken}, {Rowell}, {Sahakian}, {Samorski}, {Schilling}, {Siems},
  {Sobzynska}, {Stamm}, {Tluczykont}, {Vitale}, {V{\"o}lk}, {Wiedner}, \&
  {Wittek}}]{aharonian04}
{Aharonian}, F.~A., {Akhperjanian}, A., {Beilicke}, M., {et~al.} 2004, \apj,
  614, 897

\bibitem[{{Aharonian} {et~al.}(2006{\natexlab{d}}){Aharonian}, {Akhperjanian},
  {Bazer-Bachi}, {Beilicke}, {Benbow}, {Berge}, {Bernl{\"o}hr}, {Boisson},
  {Bolz}, {Borrel}, {Braun}, {Breitling}, {Brown}, {B{\"u}hler},
  {B{\"u}sching}, {Carrigan}, {Chadwick}, {Chounet}, {Cornils}, {Costamante},
  {Degrange}, {Dickinson}, {Djannati-Ata{\"i}}, {O'C.~Drury}, {Dubus},
  {Egberts}, {Emmanoulopoulos}, {Espigat}, {Feinstein}, {Ferrero}, {Fiasson},
  {Fontaine}, {Funk}, {Funk}, {Gallant}, {Giebels}, {Glicenstein}, {Goret},
  {Hadjichristidis}, {Hauser}, {Hauser}, {Heinzelmann}, {Henri}, {Hermann},
  {Hinton}, {Hofmann}, {Holleran}, {Horns}, {Jacholkowska}, {de Jager},
  {Kh{\'e}lifi}, {Komin}, {Konopelko}, {Kosack}, {Latham}, {Le Gallou},
  {Lemi{\`e}re}, {Lemoine-Goumard}, {Lohse}, {Martin}, {Martineau-Huynh},
  {Marcowith}, {Masterson}, {McComb}, {de Naurois}, {Nedbal}, {Nolan},
  {Noutsos}, {Orford}, {Osborne}, {Ouchrif}, {Panter}, {Pelletier}, {Pita},
  {P{\"u}hlhofer}, {Punch}, {Raubenheimer}, {Raue}, {Rayner}, {Reimer},
  {Reimer}, {Ripken}, {Rob}, {Rolland}, {Rowell}, {Sahakian}, {Saug{\'e}},
  {Schlenker}, {Schlickeiser}, {Schwanke}, {Sol}, {Spangler}, {Spanier},
  {Steenkamp}, {Stegmann}, {Superina}, {Tavernet}, {Terrier}, {Th{\'e}oret},
  {Tluczykont}, {van Eldik}, {Vasileiadis}, {Venter}, {Vincent}, {V{\"o}lk},
  {Wagner}, \& {Ward}}]{aharonian06}
{Aharonian}, F.~A., {Akhperjanian}, A.~G., {Bazer-Bachi}, A.~R., {et~al.}
  2006{\natexlab{d}}, \aap, 457, 899

\bibitem[{{Albert} {et~al.}(2008){Albert}, {Aliu}, {Anderhub}, {Antoranz},
  {Armada}, {Baixeras}, {Barrio}, {Bartko}, {Bastieri}, {Becker}, {Bednarek},
  {Berger}, {Bigongiari}, {Biland}, {Bock}, {Bordas}, {Bosch-Ramon}, {Bretz},
  {Britvitch}, {Camara}, {Carmona}, {Chilingarian}, {Coarasa}, {Commichau},
  {Contreras}, {Cortina}, {Costado}, {Curtef}, {Danielyan}, {Dazzi}, {De
  Angelis}, {Delgado}, {de los Reyes}, {De Lotto}, {Domingo-Santamar{\'{\i}}a},
  {Dorner}, {Doro}, {Errando}, {Fagiolini}, {Ferenc}, {Fern{\'a}ndez}, {Firpo},
  {Flix}, {Fonseca}, {Font}, {Fuchs}, {Galante}, {Garc{\'{\i}}a-L{\'o}pez},
  {Garczarczyk}, {Gaug}, {Giller}, {Goebel}, {Hakobyan}, {Hayashida},
  {Hengstebeck}, {Herrero}, {H{\"o}hne}, {Hose}, {Hsu}, {Jacon}, {Jogler},
  {Kosyra}, {Kranich}, {Kritzer}, {Laille}, {Lindfors}, {Lombardi}, {Longo},
  {L{\'o}pez}, {L{\'o}pez}, {Lorenz}, {Majumdar}, {Maneva}, {Mannheim},
  {Mansutti}, {Mariotti}, {Mart{\'{\i}}nez}, {Mazin}, {Merck}, {Meucci},
  {Meyer}, {Miranda}, {Mirzoyan}, {Mizobuchi}, {Moralejo}, {Nieto}, {Nilsson},
  {Ninkovic}, {O{\~n}a-Wilhelmi}, {Otte}, {Oya}, {Paneque}, {Panniello},
  {Paoletti}, {Paredes}, {Pasanen}, {Pascoli}, {Pauss}, {Pegna}, {Persic},
  {Peruzzo}, {Piccioli}, {Poller}, {Prandini}, {Puchades}, {Raymers}, {Rhode},
  {Rib{\'o}}, {Rico}, {Rissi}, {Robert}, {R{\"u}gamer}, {Saggion},
  {S{\'a}nchez}, {Sartori}, {Scalzotto}, {Scapin}, {Schmitt}, {Schweizer},
  {Shayduk}, {Shinozaki}, {Shore}, {Sidro}, {Sillanp{\"a}{\"a}}, {Sobczynska},
  {Stamerra}, {Stark}, {Takalo}, {Temnikov}, {Tescaro}, {Teshima}, {Tonello},
  {Torres}, {Turini}, {Vankov}, {Vitale}, {Wagner}, {Wibig}, {Wittek},
  {Zandanel}, {Zanin}, \& {Zapatero}}]{albert08}
{Albert}, J., {Aliu}, E., {Anderhub}, H., {et~al.} 2008, \apj, 674, 1037

\bibitem[{{Amato} {et~al.}(2003){Amato}, {Guetta}, \& {Blasi}}]{amato03}
{Amato}, E., {Guetta}, D., \& {Blasi}, P. 2003, \aap, 402, 827

\bibitem[{{Atoyan} \& {Aharonian}(1996)}]{atoyan96}
{Atoyan}, A.~M. \& {Aharonian}, F.~A. 1996, \mnras, 278, 525

\bibitem[{{Baars} \& {Hartsuijker}(1972)}]{baars72}
{Baars}, J.~W.~M. \& {Hartsuijker}, A.~P. 1972, \aap, 17, 172

\bibitem[{{Bandiera} {et~al.}(2002){Bandiera}, {Neri}, \&
  {Cesaroni}}]{bandiera02}
{Bandiera}, R., {Neri}, R., \& {Cesaroni}, R. 2002, \aap, 386, 1044

\bibitem[{{Bednarek} \& {Bartosik}(2003)}]{bednarek03}
{Bednarek}, W. \& {Bartosik}, M. 2003, \aap, 405, 689

\bibitem[{{Begelman}(1999)}]{begelman99}
{Begelman}, M.~C. 1999, \apj, 512, 755

\bibitem[{{Blumenthal} \& {Gould}(1970)}]{blumenthal70}
{Blumenthal}, G.~R. \& {Gould}, R.~J. 1970, Reviews of Modern Physics, 42, 237

\bibitem[{{Bogovalov} {et~al.}(2005){Bogovalov}, {Chechetkin}, {Koldoba}, \&
  {Ustyugova}}]{bogovalov05}
{Bogovalov}, S.~V., {Chechetkin}, V.~M., {Koldoba}, A.~V., \& {Ustyugova},
  G.~V. 2005, \mnras, 358, 705

\bibitem[{{Bogovalov} \& {Khangoulian}(2002)}]{bogovalov02}
{Bogovalov}, S.~V. \& {Khangoulian}, D.~V. 2002, \mnras, 336, 53

\bibitem[{{Bucciantini}(2006)}]{bucciantini06a}
{Bucciantini}, N. 2006, in COSPAR, Plenary Meeting, Vol.~36, 36th COSPAR
  Scientific Assembly, 191--+

\bibitem[{{Bucciantini}(2008)}]{bucciantini08a}
{Bucciantini}, N. 2008, in American Institute of Physics Conference Series,
  Vol. 983, American Institute of Physics Conference Series, 186--194

\bibitem[{{Bucciantini} \& {Del Zanna}(2006)}]{bucciantini06}
{Bucciantini}, N. \& {Del Zanna}, L. 2006, \aap, 454, 393

\bibitem[{{Bucciantini} {et~al.}(2008){Bucciantini}, {Quataert}, {Arons},
  {Metzger}, \& {Thompson}}]{bucciantini08}
{Bucciantini}, N., {Quataert}, E., {Arons}, J., {Metzger}, B.~D., \&
  {Thompson}, T.~A. 2008, \mnras, 383, L25

\bibitem[{{De Jager} \& {Harding}(1992)}]{dejager92}
{De Jager}, O.~C. \& {Harding}, A.~K. 1992, \apj, 396, 161

\bibitem[{{De Jager} {et~al.}(1996){De Jager}, {Harding}, {Michelson}, {Nel},
  {Nolan}, {Sreekumar}, \& {Thompson}}]{dejager96}
{De Jager}, O.~C., {Harding}, A.~K., {Michelson}, P.~F., {et~al.} 1996, \apj,
  457, 253

\bibitem[{{Del Zanna} {et~al.}(2004){Del Zanna}, {Amato}, \&
  {Bucciantini}}]{delzanna04}
{Del Zanna}, L., {Amato}, E., \& {Bucciantini}, N. 2004, \aap, 421, 1063

\bibitem[{{Del Zanna} {et~al.}(2006){Del Zanna}, {Volpi}, {Amato}, \&
  {Bucciantini}}]{delzanna06}
{Del Zanna}, L., {Volpi}, D., {Amato}, E., \& {Bucciantini}, N. 2006, \aap,
  453, 621

\bibitem[{{Douvion} {et~al.}(2001){Douvion}, {Lagage}, {Cesarsky}, \&
  {Dwek}}]{douvion01}
{Douvion}, T., {Lagage}, P.~O., {Cesarsky}, C.~J., \& {Dwek}, E. 2001, \aap,
  373, 281

\bibitem[{{Enomoto} {et~al.}(2006){Enomoto}, {Tsuchiya}, {Adachi}, {Kabuki},
  {Edwards}, {Asahara}, {Bicknell}, {Clay}, {Doi}, {Gunji}, {Hara}, {Hara},
  {Hattori}, {Hayashi}, {Higashi}, {Inoue}, {Itoh}, {Kajino}, {Katagiri},
  {Kawachi}, {Kawasaki}, {Kifune}, {Kiuchi}, {Konno}, {Ksenofontov}, {Kubo},
  {Kushida}, {Matsubara}, {Mizumoto}, {Mori}, {Muraishi}, {Muraki}, {Naito},
  {Nakamori}, {Nishida}, {Nishijima}, {Ohishi}, {Patterson}, {Protheroe},
  {Sakamoto}, {Sato}, {Suzuki}, {Suzuki}, {Swaby}, {Tanimori}, {Tanimura},
  {Thornton}, {Watanabe}, {Yamaoka}, {Yamazaki}, {Yanagita}, {Yoshida},
  {Yoshikoshi}, {Yuasa}, \& {Yukawa}}]{enomoto06}
{Enomoto}, R., {Tsuchiya}, K., {Adachi}, Y., {et~al.} 2006, \apj, 638, 397

\bibitem[{{Gaensler} {et~al.}(2002){Gaensler}, {Arons}, \&
  {Kaspi}}]{gaensler02}
{Gaensler}, B.~M., {Arons}, J., \& {Kaspi}, V.~M. 2002, \apj, 569, 878

\bibitem[{{Gaensler} {et~al.}(2001){Gaensler}, {Pivovaroff}, \&
  {Garmire}}]{gaensler01}
{Gaensler}, B.~M., {Pivovaroff}, M.~J., \& {Garmire}, G.~P. 2001, \apjl, 556,
  L107

\bibitem[{{Gaensler} \& {Slane}(2006)}]{gaensler06}
{Gaensler}, B.~M. \& {Slane}, P.~O. 2006, \araa, 44, 17

\bibitem[{{Gallant}(2007)}]{gallant07}
{Gallant}, Y.~A. 2007, \apss, 309, 197

\bibitem[{{Gould}(1965)}]{gould65}
{Gould}, R.~J. 1965, Physical Review Letters, 15, 577

\bibitem[{{Gould}(1979)}]{gould79}
{Gould}, R.~J. 1979, \aap, 76, 306

\bibitem[{{Green} {et~al.}(2004){Green}, {Tuffs}, \& {Popescu}}]{green04}
{Green}, D.~A., {Tuffs}, R.~J., \& {Popescu}, C.~C. 2004, \mnras, 355, 1315

\bibitem[{{H.~E.~S.~S.~Collaboration: A.~Djannati-Atai}
  {et~al.}(2007){H.~E.~S.~S.~Collaboration: A.~Djannati-Atai}, {De Jager},
  {Terrier}, {Gallant}, \& {Hoppe}}]{djannati07}
{H.~E.~S.~S.~Collaboration: A.~Djannati-Atai}, {De Jager}, O.~C., {Terrier},
  R., {Gallant}, Y.~A., \& {Hoppe}, S. 2007, astro-ph/0710.2247

\bibitem[{{Helfand} {et~al.}(2001){Helfand}, {Gotthelf}, \&
  {Halpern}}]{helfand01}
{Helfand}, D.~J., {Gotthelf}, E.~V., \& {Halpern}, J.~P. 2001, \apj, 556, 380

\bibitem[{{Hennessy} {et~al.}(1992){Hennessy}, {O'Connell}, {Cheng}, {Bohlin},
  {Collins}, {Gull}, {Hintzen}, {Isensee}, {Landsman}, {Roberts}, {Smith},
  {Smith}, \& {Stecher}}]{hennessy92}
{Hennessy}, G.~S., {O'Connell}, R.~W., {Cheng}, K.~P., {et~al.} 1992, \apjl,
  395, L13

\bibitem[{{Hester} {et~al.}(2002){Hester}, {Mori}, {Burrows}, {Gallagher},
  {Graham}, {Halverson}, {Kader}, {Michel}, \& {Scowen}}]{hester02}
{Hester}, J.~J., {Mori}, K., {Burrows}, D., {et~al.} 2002, \apjl, 577, L49

\bibitem[{{Jones}(1968)}]{jones68}
{Jones}, F.~C. 1968, \prd, 167, 1159

\bibitem[{{Kargaltsev} \& {Pavlov}(2008)}]{pavlov08}
{Kargaltsev}, O. \& {Pavlov}, G.~G. 2008, in American Institute of Physics
  Conference Series, Vol. 983, American Institute of Physics Conference Series,
  171--185

\bibitem[{{Kennel} \& {Coroniti}(1984{\natexlab{a}})}]{kennel84a}
{Kennel}, C.~F. \& {Coroniti}, F.~V. 1984{\natexlab{a}}, \apj, 283, 694

\bibitem[{{Kennel} \& {Coroniti}(1984{\natexlab{b}})}]{kennel84b}
{Kennel}, C.~F. \& {Coroniti}, F.~V. 1984{\natexlab{b}}, \apj, 283, 710

\bibitem[{{Kirk} {et~al.}(2007){Kirk}, {Lyubarsky}, \& {Petri}}]{kirk07}
{Kirk}, J.~G., {Lyubarsky}, Y., \& {Petri}, J. 2007, astro-ph/0703116

\bibitem[{{Komissarov} \& {Lyubarsky}(2003)}]{komissarov03}
{Komissarov}, S.~S. \& {Lyubarsky}, Y.~E. 2003, \mnras, 344, L93

\bibitem[{{Komissarov} \& {Lyubarsky}(2004)}]{komissarov04}
{Komissarov}, S.~S. \& {Lyubarsky}, Y.~E. 2004, \mnras, 349, 779

\bibitem[{{Kuiper} {et~al.}(2001){Kuiper}, {Hermsen}, {Cusumano}, {Diehl},
  {Sch{\"o}nfelder}, {Strong}, {Bennett}, \& {McConnell}}]{kuiper01}
{Kuiper}, L., {Hermsen}, W., {Cusumano}, G., {et~al.} 2001, \aap, 378, 918

\bibitem[{{Ling} \& {Wheaton}(2003)}]{ling03}
{Ling}, J.~C. \& {Wheaton}, W.~A. 2003, \apj, 598, 334

\bibitem[{{Lu} {et~al.}(2002){Lu}, {Wang}, {Aschenbach}, {Durouchoux}, \&
  {Song}}]{lu02}
{Lu}, F.~J., {Wang}, Q.~D., {Aschenbach}, B., {Durouchoux}, P., \& {Song},
  L.~M. 2002, \apjl, 568, L49

\bibitem[{{Lyubarsky}(2002)}]{lyubarsky02}
{Lyubarsky}, Y.~E. 2002, \mnras, 329, L34

\bibitem[{{Marsden} {et~al.}(1984){Marsden}, {Gillet}, {Jennings}, {Emerson},
  {de Jong}, \& {Olnon}}]{marsden84}
{Marsden}, P.~L., {Gillet}, F.~C., {Jennings}, R.~E., {et~al.} 1984, \apjl,
  278, L29

\bibitem[{{Masterson} {et~al.}(2005){Masterson}, {Benbow}, \& {Van
  Eldik}}]{masterson05}
{Masterson}, C., {Benbow}, W.~R., \& {Van Eldik}. 2005, in International Cosmic
  Ray Conference, Vol.~4, International Cosmic Ray Conference, 143--146

\bibitem[{{Melatos} {et~al.}(2005){Melatos}, {Scheltus}, {Whiting},
  {Eikenberry}, {Romani}, {Rigaut}, {Spitkovsky}, {Arons}, \&
  {Payne}}]{melatos05}
{Melatos}, A., {Scheltus}, D., {Whiting}, M.~T., {et~al.} 2005, \apj, 633, 931

\bibitem[{{Mezger} {et~al.}(1986){Mezger}, {Tuffs}, {Chini}, {Kreysa}, \&
  {Gemuend}}]{mezger86}
{Mezger}, P.~G., {Tuffs}, R.~J., {Chini}, R., {Kreysa}, E., \& {Gemuend}, H.-P.
  1986, \aap, 167, 145

\bibitem[{{Mori} {et~al.}(2004){Mori}, {Burrows}, {Hester}, {Pavlov},
  {Shibata}, \& {Tsunemi}}]{mori04}
{Mori}, K., {Burrows}, D.~N., {Hester}, J.~J., {et~al.} 2004, \apj, 609, 186

\bibitem[{{Much} {et~al.}(1995){Much}, {Bennett}, {Buccheri}, {Busetta},
  {Diehl}, {Forrest}, {Hermsen}, {Kuiper}, {Lichti}, {McConnell}, {Ryan},
  {Schoenfelder}, {Steinle}, {Strong}, \& {Varendorff}}]{much95}
{Much}, R., {Bennett}, K., {Buccheri}, R., {et~al.} 1995, \aap, 299, 435

\bibitem[{{Nolan} {et~al.}(1993){Nolan}, {Arzoumanian}, {Bertsch}, {Chiang},
  {Fichtel}, {Fierro}, {Hartman}, {Hunter}, {Kanbach}, {Kniffen}, {Kwok},
  {Lin}, {Mattox}, {Mayer-Hasselwander}, {Michelson}, {von Montigny}, {Nel},
  {Nice}, {Pinkau}, {Rothermel}, {Schneid}, {Sommer}, {Sreekumar}, {Taylor}, \&
  {Thompson}}]{nolan93}
{Nolan}, P.~L., {Arzoumanian}, Z., {Bertsch}, D.~L., {et~al.} 1993, \apj, 409,
  697

\bibitem[{{Pavlov} {et~al.}(2003){Pavlov}, {Teter}, {Kargaltsev}, \&
  {Sanwal}}]{pavlov03}
{Pavlov}, G.~G., {Teter}, M.~A., {Kargaltsev}, O., \& {Sanwal}, D. 2003, \apj,
  591, 1157

\bibitem[{{Rybicki} \& {Lightman}(1979)}]{rybicki79}
{Rybicki}, G.~B. \& {Lightman}, A.~P. 1979, {Radiative processes in
  astrophysics} (New York: Wiley)

\bibitem[{{Slane} {et~al.}(2004){Slane}, {Helfand}, {van der Swaluw}, \&
  {Murray}}]{slane04}
{Slane}, P., {Helfand}, D.~J., {van der Swaluw}, E., \& {Murray}, S.~S. 2004,
  \apj, 616, 403

\bibitem[{{Spitkovsky} \& {Arons}(2004)}]{spitkovsky04}
{Spitkovsky}, A. \& {Arons}, J. 2004, \apj, 603, 669

\bibitem[{{Strom} \& {Greidanus}(1992)}]{strom92}
{Strom}, R.~G. \& {Greidanus}, H. 1992, \nat, 358, 654

\bibitem[{{V\'eron-cetty} \& {Woltjer}(1993)}]{veron93}
{V\'eron-cetty}, M.~P. \& {Woltjer}, L. 1993, \aap, 270, 370

\bibitem[{{Weisskopf} {et~al.}(2000){Weisskopf}, {Hester}, {Tennant}, {Elsner},
  {Schulz}, {Marshall}, {Karovska}, {Nichols}, {Swartz}, {Kolodziejczak}, \&
  {O'Dell}}]{weisskopf00}
{Weisskopf}, M.~C., {Hester}, J.~J., {Tennant}, A.~F., {et~al.} 2000, \apjl,
  536, L81

\end{thebibliography}
\end{document}